\documentclass[acmtog,screen,authorversion,nonacm]{acmart}

\usepackage{booktabs}
\setcitestyle{numbers}

\usepackage[ruled]{algorithm2e} 

\SetAlFnt{\small}
\SetAlCapFnt{\small}
\SetAlCapNameFnt{\small}
\SetAlCapHSkip{0pt}
\usepackage{comment}
\usepackage{wrapfig}
\usepackage{graphicx}
\usepackage{subcaption}
\usepackage{tabularx}
\usepackage{multirow}
\usepackage{algpseudocode}
\usepackage{cleveref}
\usepackage{amsmath}
\usepackage{caption}
\usepackage{tikz}
\usetikzlibrary{spy,calc}

\acmJournal{TOG}

\begin{teaserfigure}
  \centering
  \includegraphics[width=0.99\textwidth]{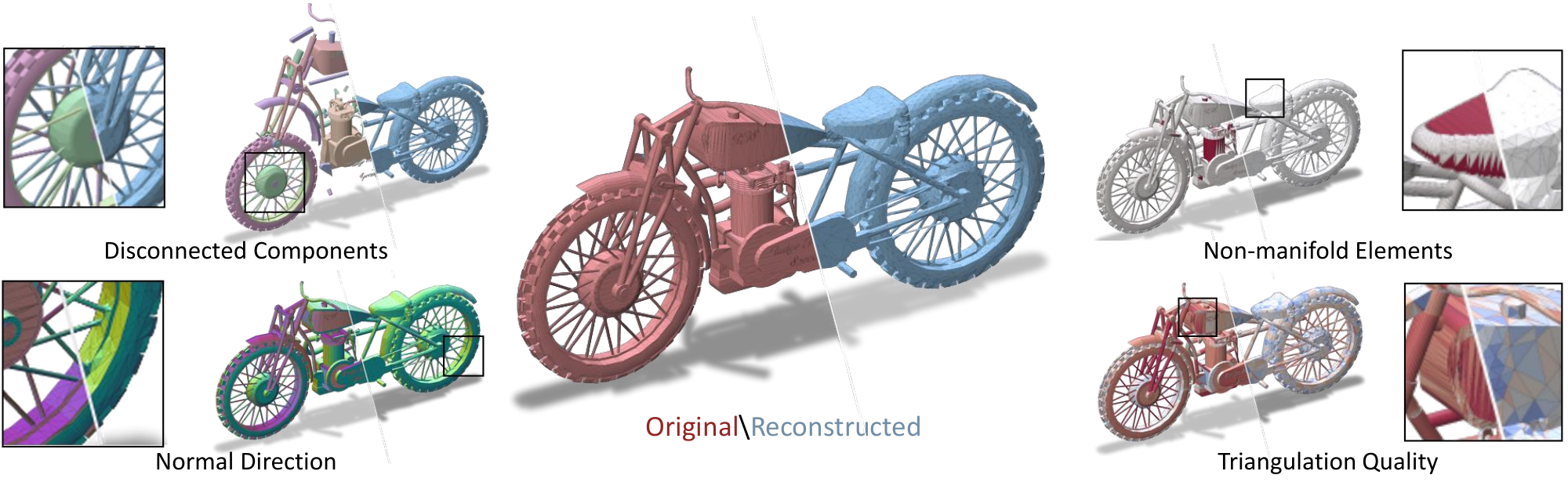}
  \caption{A triangle soup with $\sim$ 240k faces is reconstructed using our method. Our reconstructions preserve the input geometry well (middle) using a single connected component (top left, each component is colored differently) that is guaranteed to be 2-manifold (top right, non-manifold elements in red), with well-oriented normal orientations (bottom left) and high triangulation quality (bottom right, blue is better).}
  \label{fig:teaser}
\end{teaserfigure}

\begin{document}
\title{ROAR: Robust Adaptive Reconstruction of Shapes Using Planar Projections}

\author{Amir Barda}
\authornotemark[1]{}
\affiliation{%
  \institution{Tel Aviv University}
  \city{Tel Aviv}
  \country{Israel}}
\email{amirbarda@mail.tau.ac.il}
\author{Yotam Erel}
\authornotemark[1]{}
\affiliation{%
  \institution{Tel Aviv University}
  \city{Tel Aviv}
  \country{Israel}
}
\email{erelyotam@gmail.com}
\author{Yoni Kasten}
\affiliation{%
  \institution{Nvidia Research}
  \city{Tel Aviv}
  \country{Israel}
}
\email{yonikasten@gmail.com}
\author{Amit H. Bermano}
\affiliation{%
  \institution{Tel Aviv University}
  \city{Tel Aviv}
  \country{Israel}
}
\email{amit.bermano@gmail.com}

\begin{abstract}
\renewcommand{\thefootnote}{\fnsymbol{footnote}}
\footnotetext[1]{The authors contributed equally to this work.}
The majority of existing large 3D shape datasets contain meshes that lend themselves extremely well to visual applications such as rendering, yet tend to be topologically invalid (i.e, contain non-manifold edges and vertices, disconnected components, self-intersections). Therefore, it is of no surprise that state of the art studies in shape understanding do not explicitly use this 3D information, but rather focus on rendering based approaches.

In conjunction with this, triangular meshes remain the dominant shape representation for many tasks, and their connectivity remain a relatively untapped source of potential for more profound shape reasoning.

In this paper, we introduce ROAR, an iterative geometry/topology evolution approach to reconstruct 2-manifold triangular meshes from arbitrary 3D shape representations, that is highly suitable for large existing in-the-wild datasets.

ROAR leverages the visual prior large datasets exhibit by evolving the geometry of the mesh via a 2D render loss, but still observes sub-pixel resolution features using a novel 3D projection loss, the Planar Projection. After each geometry iteration, our system performs topological corrections. Self-intersections are reduced following a geometrically motivated attenuation term, and triangular resolution is added to required regions using a novel face scoring function. These steps alternate until convergence is achieved, yielding a high-quality manifold mesh adhering to the requested triangle count budget.

We evaluate ROAR on the notoriously messy yet popular dataset ShapeNet, and present ShapeROAR --- a topologically valid yet still geometrically accurate version of ShapeNet. We compare our results to various state-of-the-art reconstruction methods and demonstrate superiority in shape faithfulness, topological correctness, and triangulation quality.
In addition, we demonstrate reconstructing a mesh from a neural Signed Distance Functions (SDF), and achieve comparable Chamfer distance with much fewer SDF sampling operations than the commonly used Marching Cubes approach.
\end{abstract}


\maketitle

\section{Introduction}

The field of machine learning has made significant progress in recent times, leading to remarkable accomplishments in the realm of 3D shape comprehension tasks, encompassing both analysis and synthesis. Since data-driven approaches mostly rely on large datasets to distill information, which tend to grow exponentially larger, the quality and nature of such datasets heavily influence the advancement of the field. Many of the popular and established large 3D asset datasets \cite{shapenet2015, modelnet, thingi10k} as well as recently released ones \cite{deitke2022objaverse}, consist of shapes with countless topological errors of all types. These include everything from non-manifoldness, through self-intersections, to shapes broken into overlapping parts and inconsistent normals (See \Cref{fig:teaser}). 

These errors greatly undermine the unlocked potential and usability of a significant part of 3D shapes and datasets available today. We argue that for these reasons, many if not all of the state of the art achievements that use these datasets relating to tasks such as classification and segmentation \cite{deitke2022objaverse} and even generation \cite{gao2022get3d} rely on 2D renders of the data instead of directly learning over the 3D shape.

To address these issues, and to further the usability of large (and messy) 3D mesh datasets, a suitable reconstruction operation is required. 
Reconstruction techniques seek a different sampling of the same underlying shape, while maximizing faithfulness and balancing out or improving some desirable properties such as triangle count and tessellation quality. 
 
In addition, most 3D processing operations, including modern learning-based techniques, expect a 2-manifold mesh as input \cite{meshcnn, pd_mesh_cnn, hodgenet} and hence topological validity is perhaps the most important aspect of reconstruction.

Reconstruction approaches employing regular volumetric sampling  (i.e. a voxel grid), excel in providing guarantees on topological validity and naturally support shapes of arbitrary genus, but they also tend to exhibit sampling artifacts and fail to express fine details or large but thin parts due to resolution constraints (\cite{manifold, manifold_plus, tetwild, ftetwild}). Surface-based approaches, on the other hand, that evolve an initial shape or directly manipulate the target, are more expressive in this sense \cite{competing_fronts, advancing_fronts_cgal, MeshRepair}. However, as it turns out, such approaches require careful selection of hyper parameters per target, and in practice tend to be too slow to operate over large scale datasets due to global optimization steps. Recent appearance based approaches \cite{palfinger2022continuous, large_steps}, on the other hand, have shown that using differentiable 2D renderings of the source and target shapes is a powerful tool for shape reconstruction, offering better convergence. The 2D nature of this approach, however, can yield undesirable artifacts if not properly accounted for as demonstrated by \citet{large_steps}. Furthermore, optimizing fine features on the surface using the render based loss relies on vertex normals, which are of poor quality or do not exist at all in many practical cases. In addition, features with sub-pixel resolution (such as corners) are disregarded in the optimization (\Cref{tab:ablations}). Lastly, many such techniques are prone to get stuck in local minima, critically damaging tessellation quality, especially if using loss terms involving global distances (e.g. Chamfer distance, see \Cref{fig:projection_loss}). All these drawbacks are the reason such methods have not been used for the purpose of reconstructing and repairing large datasets as of today.

In this paper, we present ROAR --- a practical and robust reconstruction solution accurate and reliable enough for complete dataset repair. Our proposed technique successfully retains topological correctness, produces high-quality triangulation while preserving faithfulness, and is implemented completely using an auto-diff library (PyTorch \cite{pytorch}) on the GPU.

Following recent literature~\cite{large_steps, palfinger2022continuous}, our iterative approach evolves an initial shape by constantly correcting the mesh topology after every geometry update. This ensures mesh validity, self-intersection minimization, and efficient triangle allocation. The main challenge of taking this approach is given a triangle count budget to both assign enough resolution to detailed regions and to remove problematic triangles, such as those that cause local self-intersections. 

After a pre-processing step, we first extract an initial mesh using an off-the-shelf reconstruction approach. Then, our method refines the proposed solution by alternating between geometric changes and topological corrections. The geometry is governed by both a 3D planar projection term \cite{planar_icp} used as a novel loss function that better preserves geometric details, and a 2D image loss term that regularizes convergence both globally and locally. Each geometric iteration is followed by an edge collapse operation that prevents self-intersections from persisting and evolving. We then continue to add resolution to the mesh adaptively, using a novel expansion of a rapid self-intersection estimator \cite{palfinger2022continuous}.

We implement a prototype for ROAR using a GPU, making computation an order of magnitude faster than on the CPU, operations feasible, and the algorithm simple to integrate. We evaluate ROAR on reconstruction tasks, showing better reconstruction performance for a large scale triangle soup dataset \cite{shapenet2015}. This includes consistent topological validity, better triangle quality, and better shape preservation compared to the state-of-the-art. In addition, we demonstrate reconstructing a mesh directly from implicit neural SDF (signed distance function), and show that our approach can express the same level of detail that uniform 3D sampling-based approaches (e.g., marching cubes) achieve, with significantly less sampling operations.

Lastly, we present ShapeROAR, a topologically valid yet still geometrically accurate version of the ShapeNet dataset \cite{shapenet2015}, rendering learning over shapes a simpler challenge.

In summary, the core contributions of this paper are:
\begin{itemize}
    \item A novel reconstruction pipeline, adaptively allocating triangle resolution in required regions while maintaining topological validity and tessellation quality.
    \item ShapeROAR, a topologically valid reconstruction of ShapeNet.
    \item A planar projection operation formulated as a novel loss term, substituting shading gradients when unavailable (e.g. triangle soups).
    \item A novel face score criterion for rapidly detecting local shape variation in-situ.
\end{itemize}

An open sourced implementation and the cleaned dataset ShapeROAR will be made publicly available.

\section{Related Work}
We chose to focus on specific methods that relate to ROAR by employing similar techniques or intend to solve a similar problem while having an open-access implementation that can be used for comparisons. It is important to note however that despite providing proof for unconditional robustness (output is 2-manifold, with no self intersections), in practice all of the official implementations we found do not obtain this (\Cref{sec:in-the-wild}), emphasizing the gap that exists between theory and practice. Additionally, some prior studies cannot directly be applied to large scale datasets due to the sheer amount of compute time taken per-mesh. We discuss this issue in \Cref{sec:in-the-wild}.

\subsection{Surface based}
A common goal for reconstruction techniques is to resample the input domain to obtain a topologically valid triangulation while approximating the input shape as much as possible. The output of such techniques is useful for many downstream tasks such as simulations and UV texture unwrapping. 
\citet{alphawrap_cgal} iteratively construct a subcomplex of a 3D Delaunay triangulation by starting from a simple 3D Delaunay triangulation enclosing the input, and then iteratively removing eligible tetrahedra that lie on the boundary of the complex. The result is oriented, 2-manifold, and without self-intersections. More recently, \citet{MeshRepair} presents a surface-based technique to repair triangle soups operating directly on the input, by first locally repairing patches using visual cues and later globally optimizing for a manifold mesh formulated as a ILP. The main disadvantage is the extremely long running time, disallowing reconstruction for large scale datasets, and the rather low triangulation quality of results, as triangle quality was left unattended and is similar to the input.

Our reconstruction pipeline also yields an oriented 2-manifold surface, and despite not having guarantees on self-intersections their amount is low in practice due to the criterion we impose (\Cref{sec:prelim}). Our results better approximate the target for a given triangle budget, and can be obtained within a reasonable running time.

\subsection{Volume based}
Some techniques employ a change in representation approach, where the input is intersected with a voxel grid \cite{manifold, manifold_plus} and reconstructed using topological guarantees such as manifoldness and other desirable qualities. In Tet-Wild \cite{tetwild}, the authors explicitly deal with "messy" inputs by considering the input fundamentally imprecise, and performing tetrahedralization using a binary space partitioning tree of planes that wraps the input model, and later on optimizing its quality while preserving the validity. Despite not being directly used for surface reconstruction but rather volumetric mesh reconstruction, the explicit treatment of input as flawed makes this approach competitive in this task. To extract the surface, one may select the volumetric output boundary surface (i.e. faces incident to a single tetrahedron). \citet{ftetwild} extend this method by performing incremental reconstruction, and achieves better performance times while using a floating point representation. Reconstructing a surface in this manner was determined to be disadvantageous by our experiments, especially considering the computational resources associated with it (as opposed to evolving a surface mesh directly).

\subsection{Render based}
In a fairly recent line of work, differentiable rendering based techniques showed interesting new directions in evolving shapes using appearance. In \cite{nvdiffrast}, a robust and scalable neural rasterizer was developed and its application in shape reconstruction is shown over both synthetic and real data. In \cite{large_steps}, the authors develop a correction term for the render loss that allows convergence into an input shape from a sphere, with very little loss of high frequency detail that is usually associated with the usage of regularizer terms in the loss. These terms usually exist to prevent the optimized shape from folding onto itself and preventing self-intersections, due to the large and sparse gradients associated with the silhouette of the object. The authors show their geometry updates can be used with an offline reconstruction algorithm to evolve a sphere into a target mesh with intricate details. In \cite{palfinger2022continuous}, a full algorithm for reconstruction addressing topological connectivity is presented, leveraging the render loss to augment the geometry, interleaved with topological steps that assure triangles are created when necessary. Despite not using the correction term from \cite{large_steps}, results are of high quality, but they require a large number of triangles to complete the shapes and there are no topological guarantees on the output. We leverage these past findings and improve upon the quality and topological properties of the output as well as being more efficient with triangle budget by allocating triangles adaptively to surface details. Additionally, in both works, it was observed that the gradients received from the differentiable renderer can be classified into to two groups: The aforementioned sparse and strong silhouette gradients, and the dense but weaker gradients that originate from the shading, called "shading gradients", which strongly depend on well oriented normals. Both methods rely on high quality shading gradients to function well, which makes them impractical for the reconstruction of meshes with topological defects.

\subsection{Point Cloud based}
Another approach using a different intermediate representation for reconstruction is to convert the input into a pointcloud by sampling, and reconstructing the surface. Discarding connectivity information is a lossy operation, and indeed reconstructing an already connected surface is not the bread-and-butter of such applications. However, it is useful to consider them as a baseline and to draw conclusions on how effective can surface reconstruction get with versus without connectivity. One traditional technique introduced by \citet{poisson} casts the reconstruction task as a spatial Poisson problem and solves it using a linear sparse solver. This technique saw many improvements over the years, in particular, \citet{screened_poisson} incorporate the points as interpolation constraints and show improved reconstruction quality. \citet{shape_as_points} expand Poisson surface reconstruction by formulating it as a differentiable layer, enabling end-to-end optimization of watertight manifold surfaces. \citet{point2mesh} attempt to solve the same problem using a deep neural network prior and a Beam Gap loss.

\section{Method}

\begin{figure*}
    \centering
    \includegraphics[width=1.0\textwidth]{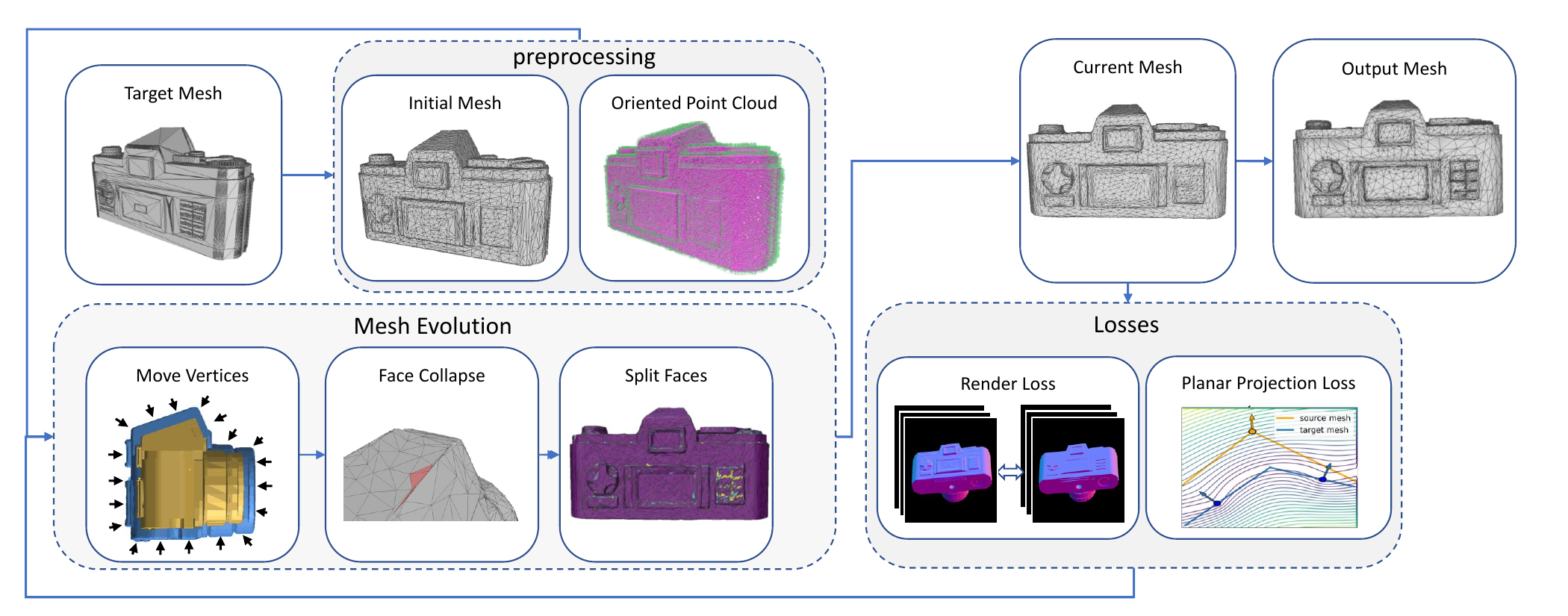}
    \caption{Mesh Reconstruction Pipeline. The target mesh, which is potentially broken topologically in several ways, is reprocessed to generate an initial source mesh. Our iterative evolution consists of three steps. The first updates vertex positions ("Move Vertices"), and the other two update connectivity by removing self-intersections as they start to form ("Face Collapse") and adding resolution to regions that require more detail ("Face Split"). Severe orientation problems with the input make any form of vertex shading challenging or even infeasible. Therefore, we flat shade using face normals, and use the planar projection loss to compensate for the now-lost shading gradients.}
    \label{fig:pipeline}
\end{figure*}

As a reminder, ROAR leverages an iterative surface evolving approach. This choice was made deliberately to allow us to exploit recent advancements with 2D differentiable rendering techniques (allowing to preserve appearance), and to also ensure topological validity (2-manifold) by choice of initial shape and operations performed on it during evolution. The main challenges associated with such an approach are efficient allocation of triangles, and maintaining local and global validity (self-intersections). Our introduced pipeline is shown in \Cref{fig:pipeline}. The inputs to our method are the raw \textit{target shape} to be reconstructed, an initial \textit{source mesh} that will evolve, and a face budget to constrain the number of triangles in the output. After a preprocess step, we evolve the state of the source mesh iteratively, where each iteration consists of three blocks. 

In the preprocessing step (\Cref{sec:preprocess_rendering}), we create the initial mesh by applying MANIFOLD ~\cite{manifold} on the original target mesh. Additionally, we generate an oriented and cleaned point cloud of the target mesh required for our loss terms.

After preprocessing, we start evolving the initial mesh: 
First, a geometry update step is performed (\Cref{sec:update_vertices}), where vertex locations are adjusted according to the loss function. To handle triangle soups, we supplement the rendering loss with a novel added term, the \textit{Planar Projection Loss} (\Cref{sec:planar_projection}), offering a robust alternative to the common Chamfer Distance loss, and acting as a replacement for shading gradients (see \Cref{sec:prelim}). We found it crucial for preserving sharp features and being robust to errors on the target. Additionally, we inhibit the creation of self-intersections by attenuating the amount of displacement a vertex can undergo according to the lengths of the edges emanating from it. 

The second step, Face Collapse, (\Cref{sec:face_collapse}) identifies and removes self-intersection as they begin to form. Self-intersections that evolve too quickly can potentially elude this process and remain uncleaned. The aforementioned attenuation factor discourages this behavior.

The next step, Face Split (\Cref{sec:face_split}), dictates where resolution should be added. To do this, we introduce the \textit{Face Score} --- an extension of the self-intersection estimator (\Cref{eq:fold_score}), that encodes local geometry changes - deciding whether triangles should be split according to the local shape of the target region they correspond to.

In the following, we start with preliminary work and continue describing the operation of each block in depth. Detailed descriptions of the hyper-parameters used are given in the supplementary material.

\subsection{Preliminary}
\label{sec:prelim}
In \citet{large_steps}, an analysis of rendering-based gradients (of image pixels with respect to vertex locations) revealed two distinct "types" of gradients that contribute to vertex movements: shading gradients, which are smooth, dense, and occur from pixels inside shape primitives, and silhouette gradients, which are sparse, strong, and occur from pixels overlapping the border of the shape. Silhouette gradients were shown to interfere greatly with shape reconstruction, specifically because they induce local self-intersections by neighboring faces folding onto themselves.
\citet{palfinger2022continuous} seek to evolve an initial sphere into a target shape using iterative geometric and connectivity updates driven by a differentiable render loss. One important insight is to continuously attempt to correct self-intersections that occur due to sparse silhouette gradients. An estimator for local self-intersections for a face $f$ was used: 
\begin{equation}
    n_N\cdot n_f < 0
    \label{eq:fold_score}
\end{equation}
 where $n_N$ is the mean of the face's vertex normals and $n_f$ is the face normal. This estimator was also discussed in previous studies \cite{manifold_plus}, though not in the context of differentiable rendering. The advantage of using such an estimator is performance, and unlike other approaches which perform complete re-triangulation of the mesh when self-intersections occur \cite{point2mesh, large_steps}, this estimator can be used to prevent self-intersections in every vertex position optimization iteration. This allows to continue operating on the same evolving surface. Another mechanism used by \citet{palfinger2022continuous} to prevent local self intersections is an attenuation factor for the vertices position updates. Namely, for a vertex position $\mathbf{x}$, being optimized with the rotation-invariant ADAM formulation ~\cite{palfinger2022continuous, ling2022vectoradam}, the update step becomes:
\begin{equation}
    \Delta \mathbf{x} = \alpha \cdot \boldsymbol{\nu} \cdot l_{ref}
    \label{eq:vertex_update_step}
\end{equation}

where $\alpha$ is the learning rate, $\boldsymbol{\nu}$ is the ADAM update rule (computed from the gradients of the loss with respect to $\mathbf{x}$) and $l_{ref}$ is a coefficient that is (initially) set to the length of the edges incident to $\mathbf{x}$ and gets updated using the average $\boldsymbol{\nu}$. In a nutshell, this further restricts the movement of the vertex, effectively reducing the amount of generated self-intersections. In addition, this term was argued to be an indicator of where new triangles are required, because $\mathbf{\nu}$, or the amount of change a vertex should undergo during the current iteration, is (indirectly) proportional to the reconstruction fidelity (as larger distances warrant larger gradients). In our experiments, we found this term to be too noisy, and propose a more geometrically motivated approach for adding resolution to the mesh.
 
\subsection{Move Vertices}
\label{sec:update_vertices}
Every iteration of the source mesh shape evolution starts by updating its geometry (vertex positions). The vertex positions are updated using a gradient-based optimizer (rotation invariant ADAM), and employ the following loss terms to drive it towards a good solution:
\begin{multline}
        L(v) = L_{Im}(v_{Source},v_{Target}) + \lambda_1 \cdot L_{Proj}(v_{Source},v_{Target}) \\+ 
    \lambda_2 \cdot L_{Proj}(v_{Target},v_{Source})
    \label{eq:loss}
\end{multline}

Where $L_{Im}$ is a 2D image loss computed over different views of the source and target mesh using a differentiable rasterizer \cite{nvdiffrast}. We use a flat shaded back-face culled rendering of the geometry with color encoding indicating face normal directions (see supplementary for examples). These appearance properties yield strong signals that sharply change with surface geometry, and are less ambiguous than shading schemes that depends on a light source, a texture or vertex normals. $L_{Proj}$ is our novel planar projection loss term that drives the geometry to better fit the target, especially in areas with pixel or sub-pixel resolution features. The projection operation driving this loss term is also used as a part of our face splitting process, and is described in further detail in \Cref{sec:planar_projection}. $\lambda_1$ and $\lambda_2$ are coefficients to balance between the terms. Additionally, we wish to constrict vertex movements to avoid self-intersections. We note that the purpose of using the attenuation factor $l_{ref}$ (\Cref{eq:vertex_update_step}), is to prevent the formation of folded faces too quickly for them to be resolved. Instead, we propose a more geometrically oriented attenuation factor, which we call $l_{att}$, for the vertex updates:
     \begin{equation}
       l_{att}(v) = min(|e|),  \forall e: v\in e
       \label{eq:l_att}
    \end{equation}
where $e$ are the incident edges. This limits the maximum displacement a vertex can undergo in a single iteration, such that faces are not allowed to flip over, making sure that self-intersections are identified. Note that $l_{att}$ must be calculated at every iteration, for all vertices. We do this by constructing a topological sparse matrix for the mesh and performing sparse-dense multiplication with the vertex tensor. Since these operations are all performed on the GPU, the run time overhead is acceptable, even when this term is calculated at every iteration. We find that our attenuation factor achieves better results (see \Cref{sec:ablations}). After the update by the optimizer, to promote isotropic triangles, a global tangential smoothing step is performed \cite{ada_remesh}.

\subsection{Face Collapse}
\label{sec:face_collapse}
During the evolution of the mesh, self-intersection may develop. We identify \textit{local} self-intersections using the folded face estimator posed in \Cref{eq:fold_score}. We resolve the found folded faces by collapsing them: we first calculate the Qslim score ~\cite{qslim} for all edges of the folded faces. Intuitively, a high Qslim score means that the edge encodes salient geometry. Our Qslim formulation also includes checks for face quality and normal flipping after collapse. We then insert these Qslim scores to a priority queue, and collapse the edges according to manifold-preserving rules \cite{hoppe}. For simplicity, we collapse the edges using the subset strategy given in ~\cite{qslim}, meaning we can only collapse an edge towards one of its end points. We end this step with a round of edge flips, to balance the vertex valences. we threshold the edge flip on the dihedral angle of the edge, in order to avoid damaging reconstruction fidelity.

\subsection{Face Split}
\label{sec:face_split}
 The Face Split block is responsible for splitting triangles in areas of interest. There is a great degree of freedom of how and where should new triangles be added. To this end, we pursue a subdivision scheme that allows us to locally increase resolution but in a parallel manner (for performance), and a score function for indicating where it should be done. The subdivision strategy we found most suitable for our purposes is the $\sqrt{3}$ subdivision \cite{sqrt3}, which inserts a vertex at the barycenter of a triangle and connects it to all its vertices. This scheme allows us to split an arbitrary subset of the triangles and maintain a valid 2-manifold triangle mesh. The ADAM parameters of the newly create barycentric vertices are simply the average of their parent vertices' ADAM parameters. In fact this plays well with general machine learning applications, where if any learnable signal is carried on the vertices of a triangle that is split, the new vertex can simply inherit the weighted average of those signals naturally. As for the determination of which faces to split, clearly, a good score function must rely both on the source mesh \textit{and} target shape. However, since no correspondence exists between them, and the target is noisy, we propose a scoring function that relies on a novel extension of the self-intersection criterion \Cref{eq:fold_score} imposed on a face:

\begin{align}
 C(f) = \begin{cases}  1-n_N \cdot n_f & \text{if } n_N\cdot n_f>0 \\
0 & \text{otherwise}. \end{cases}
     \label{eq:face_curvature}
\end{align}
    
Under this definition, when $C(f)$ nears 1, the face $f$ encodes a surface with more locally varying geometry. Note that when $n_N \cdot n_f < 0$ the face is estimated to be part of a self-intersection.
As a reminder, our goal is to score the source mesh faces for splitting. Applying the score function to the source mesh faces directly is an option, but this has two downsides: first it assumes geometry is as perfect as it can be, i.e. the source mesh vertices reached some steady-state local minima in relation to the target shape. This assumption is simply untrue in the general case (and especially initially).
\begin{wrapfigure}{r}{0.11\textwidth}
\includegraphics[width=0.10\textwidth]{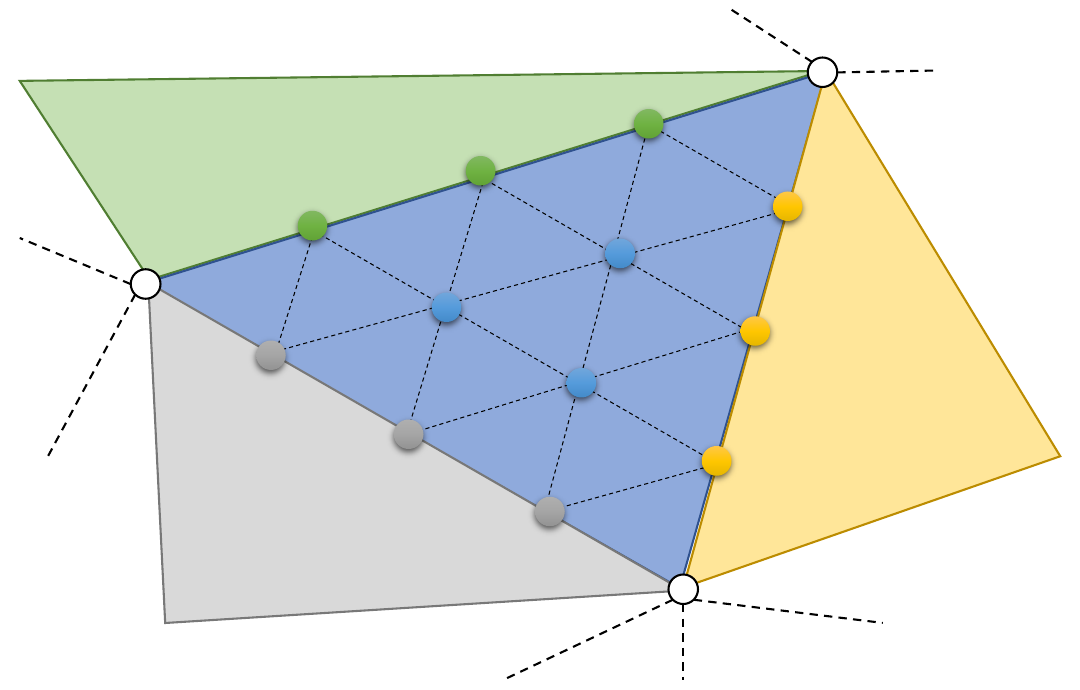}
\end{wrapfigure}
Second, the target mesh is not taken into account at all, which could be used as a guiding prior. Thus, a more elaborate application of the score function is necessary. To this end, we apply a regular face super-sampling operation on the source mesh \cite{supersample_mesh}, which creates smaller triangles all having the same area and allows fast computations of integral quantities over the parent face (see inset).

\begin{figure}[ht]
    \centering
    \begin{subfigure}{0.235\textwidth}
    \includegraphics[width=\textwidth]{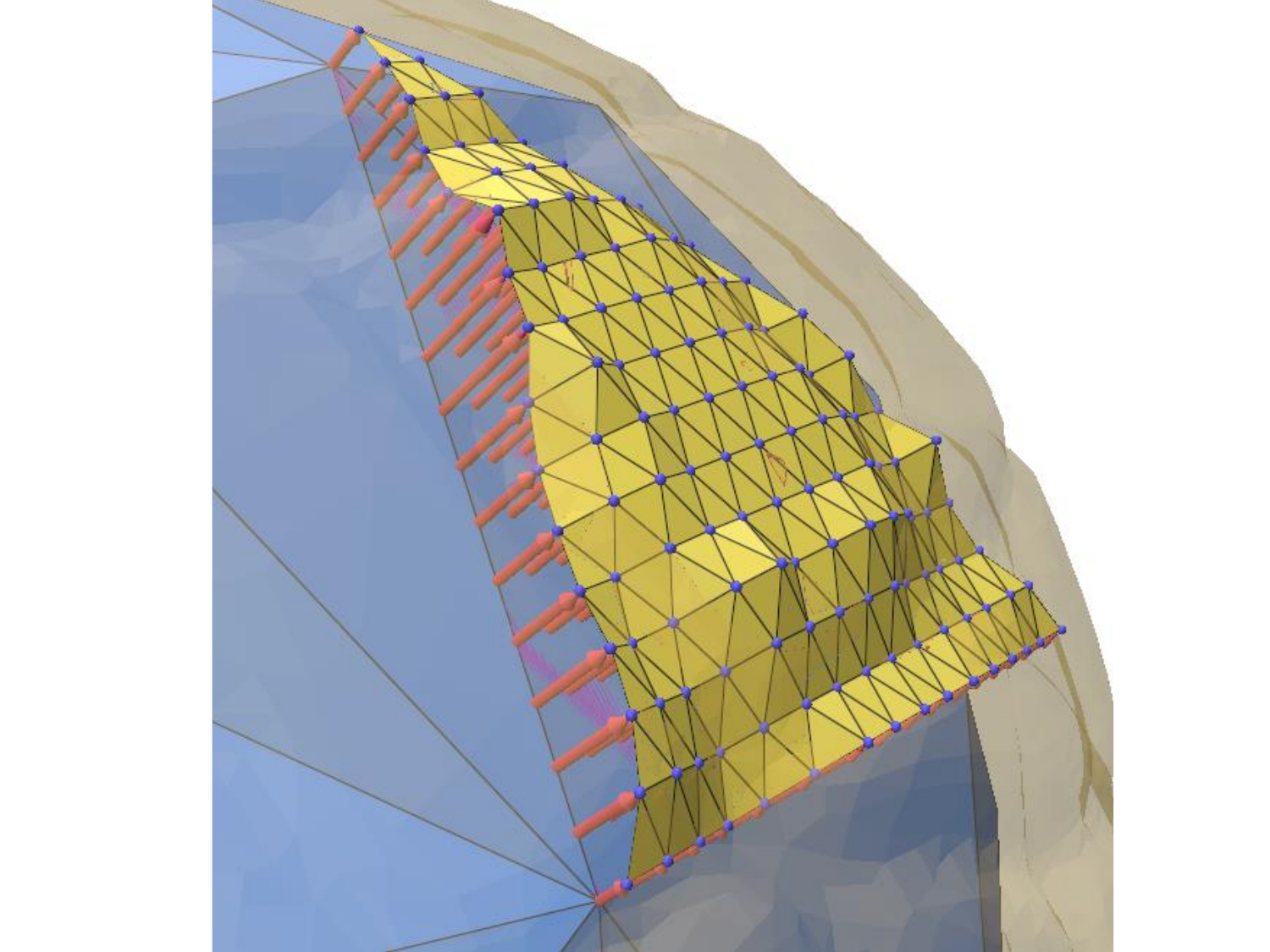}
    \caption{}
    \end{subfigure}
    \begin{subfigure}{0.235\textwidth}
    \includegraphics[width=\textwidth]{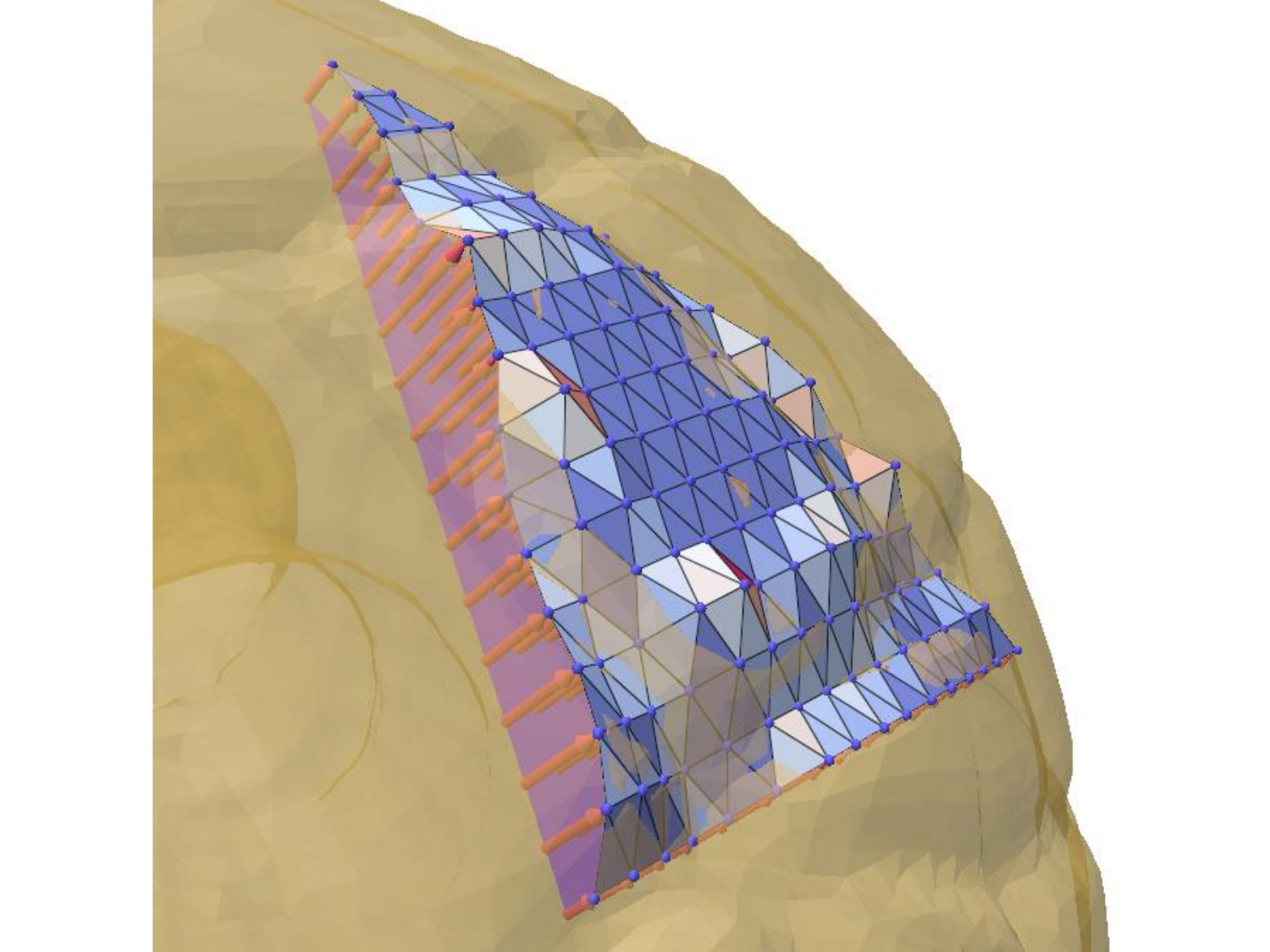}
    \caption{}
    \end{subfigure}
    \caption{Face score calculation: (a) a single supersampled face is shown on the mesh being evolved (blue), the supersampled faces are projected (red arrows) on to the target mesh (orange, transparent) using our planar projection operation. (b) $C(f)$ heatmap for the supersampled faces shown in (a), using \Cref{eq:face_curvature}. Note how the "hot" faces correspond to non-flat areas in the target mesh.}
    \label{fig:face_score}
\end{figure}

We then follow by the projection of the smaller triangles' vertices to the target mesh. By examining the resulting score of the projected super-sampled faces, we can infer the areas where more resolution is necessary on the source mesh (see \Cref{fig:face_score}). This takes into account both the source mesh current state, as well as the target shape.

The super-sampled faces' scores are then pooled to their parent faces:

\begin{equation}
     FS(f) = \sum_{i}{C(\Phi(f_i))}
     \label{eq:total_curvature}
\end{equation}

where $\Phi$ is the projection function described below.

In the final stage of this block, we promote better average vertex valence by a global manifold-persevering edge flip operation.

\subsubsection{Planar Projection}
\label{sec:planar_projection}
As mentioned, the super sampled faces' vertices are projected onto the target to approximate local geometry changes.

The natural option for performing such a projection is to find the nearest point on the mesh for every vertex of the super sampled faces. This operation however, is computationally intensive, with no real benefit over the discrete approximation counterpart (see supplementary). A common fast approximation is to fairly sample the target surface, and project a query point to its nearest sample (or the average of several nearest samples) during operation. 
 
\begin{figure}[ht]
    \centering
    \begin{subfigure}{0.235\textwidth}
    \includegraphics[width=\textwidth]{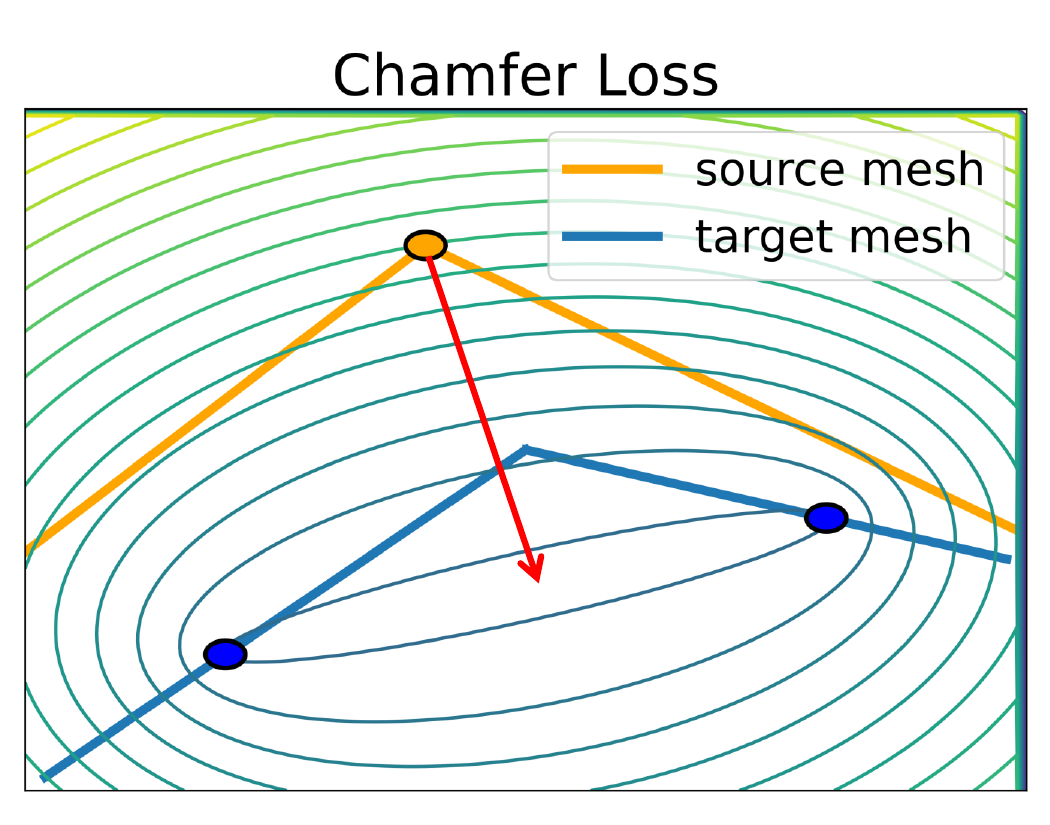}
    \caption{}
    \end{subfigure}
    \begin{subfigure}{0.235\textwidth}
    \includegraphics[width=\textwidth]{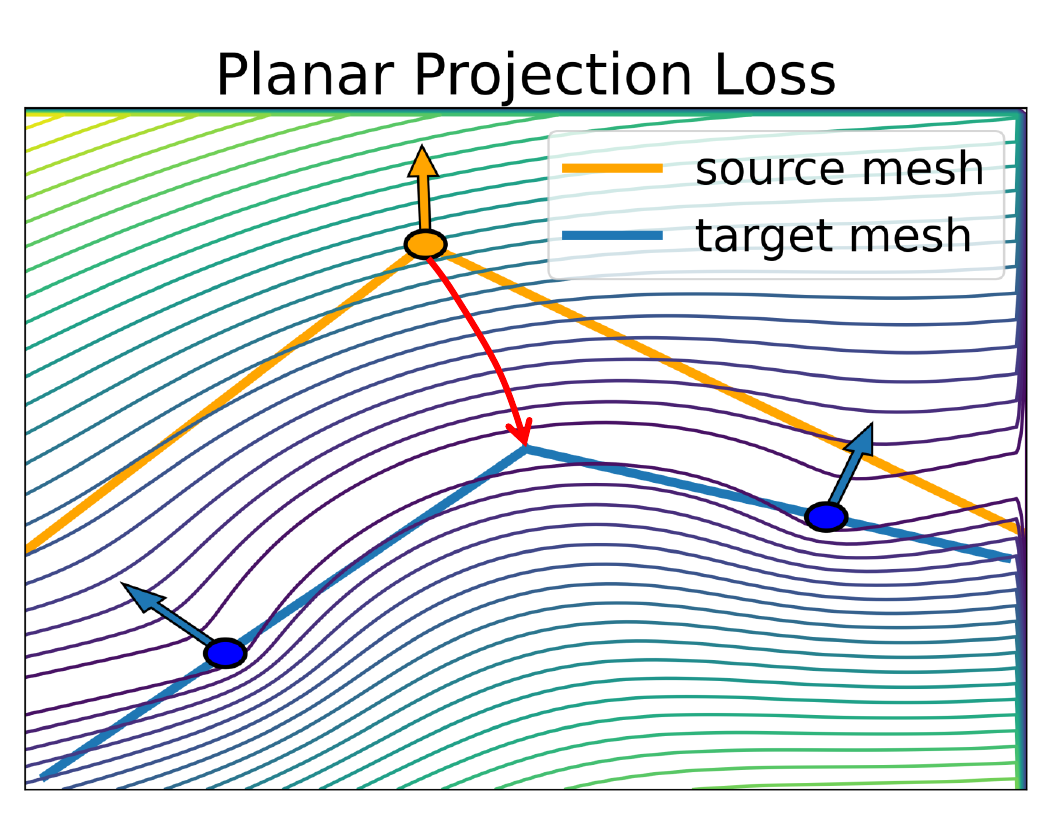}
    \caption{}
    \end{subfigure}
    \caption{Planar Projection. (a) Iso-contours of a 2D 2-nearest neighbor Chamfer loss. The minimum is achieved between the two blue sampled vertices on the target mesh, which means the corner will not be reconstructed correctly. Moreover, In the case of geometry visible by the renderer, this will not agree with the silhouette gradients, leading to sub-optimal reconstruction (b) Iso-contours of a 2D 2-nearest neighbor planar projection loss computed from \Cref{eq:projection}, with the source vertex maintaining its normal direction. The desired corner is part of the zero set, by incorporation of target normals.}
    \label{fig:pp_explain}
\end{figure}

 Unfortunately, due to the injective nature of this approximation it causes degenerated artifacts and acute overlaps especially when used as a loss term \cite{chamfer_self_intersect}, which lead to inferior surface estimation (see \Cref{fig:pp_explain}). 

 Instead, we propose two augmentations to the process; in order to preserve triangulation quality and discourage overlaps and the local minima typical to Chamfer-based projections, we project the source points only along their normal directions. This means source vertices could probably not be projected to the nearest sampled target points (or their average) exactly. We propose projecting to the \textit{plane} each sampled target point locally approximates (see \Cref{fig:pp_explain}), instead of to the point itself: 
given a point $v$ on the source mesh with normal $\hat{n}_v$, we first find its $K$ nearest neighbors on the target (and their normals $\hat{n}_{k}$) to compute a \textit{support} $s_{k}$ which is the projection of $v$ to the plane defined by each such neighbor, in the direction $\hat{n}_v$:
\begin{equation}
\begin{split}
d & = \frac{(k_i - v) \cdot \hat{n}_{k_i}}{\hat{n}_v \cdot \hat{n}_{k_i}} \\
s_{k_i} & = v + d \cdot v_n
\end{split}
\label{eq:support}
\end{equation}

The final projection point is computed as an average of all supports, weighted by their distance to $v$:

\begin{equation}
\begin{split}
w_i & = ||s_{k_i} - v||_2 \\
P(v) & = \frac{\sum_{i}{w_i \cdot s_{k_i}}}{\sum_{i}{w_i}}
\label{eq:projection}
\end{split}
\end{equation}

We call this operation \textit{Planar Projection}, and found it crucial in preserving triangulation quality, adhering well to sharp features, and being robust to errors on the target (see \Cref{fig:projection_loss}).
Additionally, note the steps to compute $P(v)$ are completely differentiable with respect to $v$. We take advantage of this in the Move Vertices step (\Cref{sec:update_vertices}), by defining a loss term $||v - P(v)||_2$ which snaps vertices into corners in areas with sub-pixel resolution features. In those areas, the render loss roughly performs random perturbations of the vertices (i.e. gradients are small, and random), but given dense enough sampling, the Planar Projection term is minimized when vertices are exactly on corners or borders (an intersection of two or more planes). A good analogy for this is quicksand: the more jiggling action occurs, the more the vertex sinks into corners. Note that the Planar Projection Loss by itself can never achieve this as $P(v)$ is always in the direction of $\hat{n}_v$, which rarely points towards the corner. See \Cref{fig:pp_explain} for a more visual explanation.

\subsection{Mesh Preprocessing}
The purpose of the preproecssing step is to create an initial mesh to begin optimization and an oriented point cloud used for our projection scheme.
\subsubsection{Rendering}
\label{sec:preprocess_rendering}
In real data, it is not uncommon to have meshes that are ill-oriented. This usually occurs due to the nature of the data formation (e.g. a designer who did not care for orientation, a reconstruction algorithm, or a result from an actual scan). Despite the fact many reconstruction algorithms treat face orientation as a feature of the input \cite{tetwild, ftetwild}, 
this poses a problem when rendering the mesh, as the renderer can wrongly render back-facing triangles, or not render front-facing triangles, causing holes and overlap artifacts to appear in the rendering. To solve this, we first remove all duplicate faces (i.e. permutations of the same vertices). We then duplicate every face but with flipped orientation - these two steps guarantee every face has exactly one twin face facing the other direction, with the visible face being rendered, and the other culled by the renderer. Additionally, we normalize the mesh by moving the average vertices' location to the origin and scaling them to a unit sphere.
\subsubsection{Mesh Initialization}
\label{sec:init_mesh}
we use MANIFOLD \cite{manifold} on the normalized target mesh, to get an initial mesh for optimization which is 2-manifold by construction. Unfortunately, MANIFOLD's output occasionally consists of non-manifold vertices. We use MeshLab's \cite{meshlab} non-manifold vertices cleaning filter to fix this by splitting these vertices. After this, the initial mesh is guaranteed to be 2-manifold.
\subsubsection{Sampling}
\label{sec:orient_target}
\begin{figure}[ht]
    \centering
    \begin{subfigure}{0.13\textwidth}
    \centering
    \includegraphics[width=0.9\textwidth]{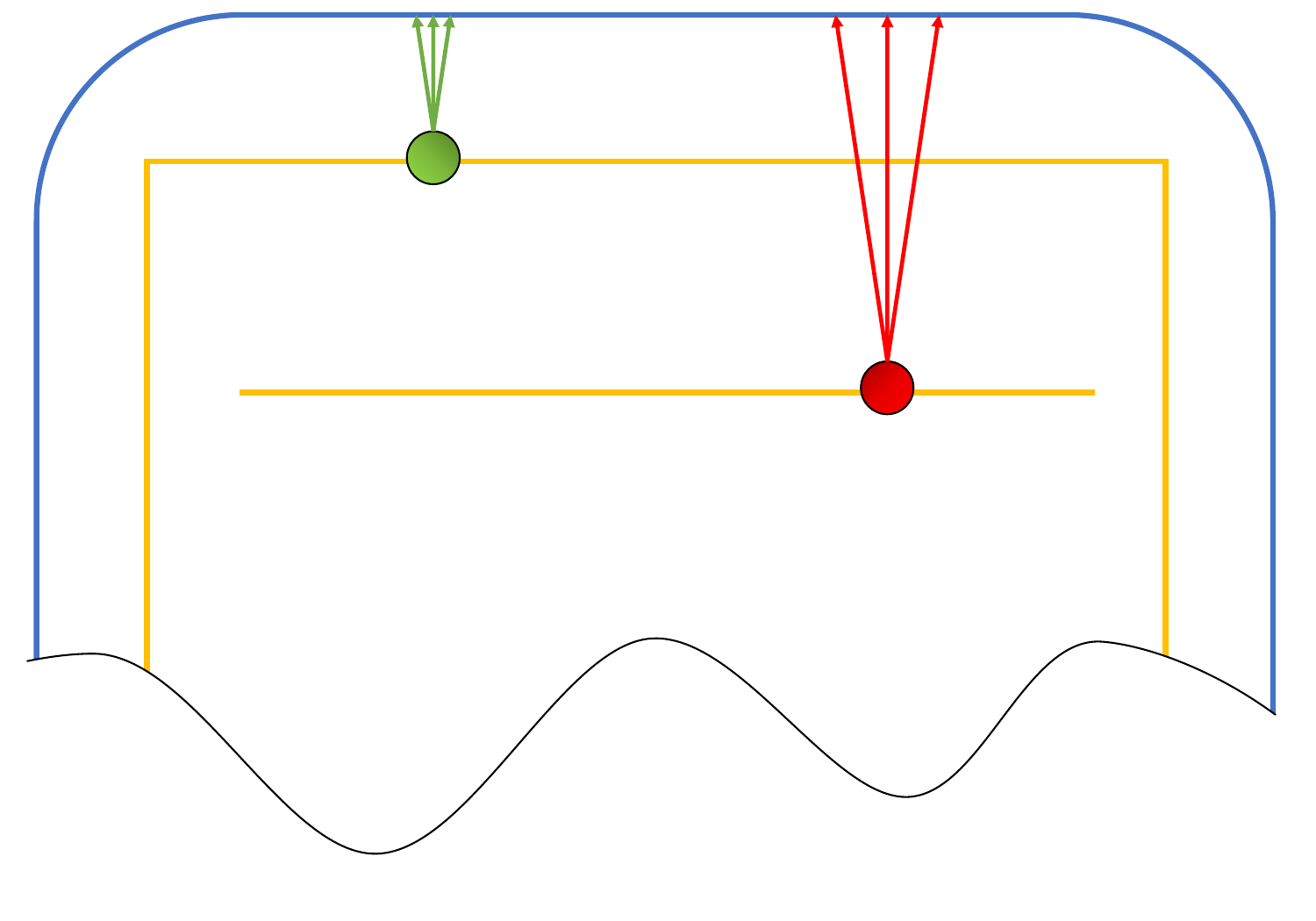}
    \end{subfigure}
    \begin{subfigure}{0.33\textwidth}
    \centering
    \includegraphics[width=0.9\textwidth]{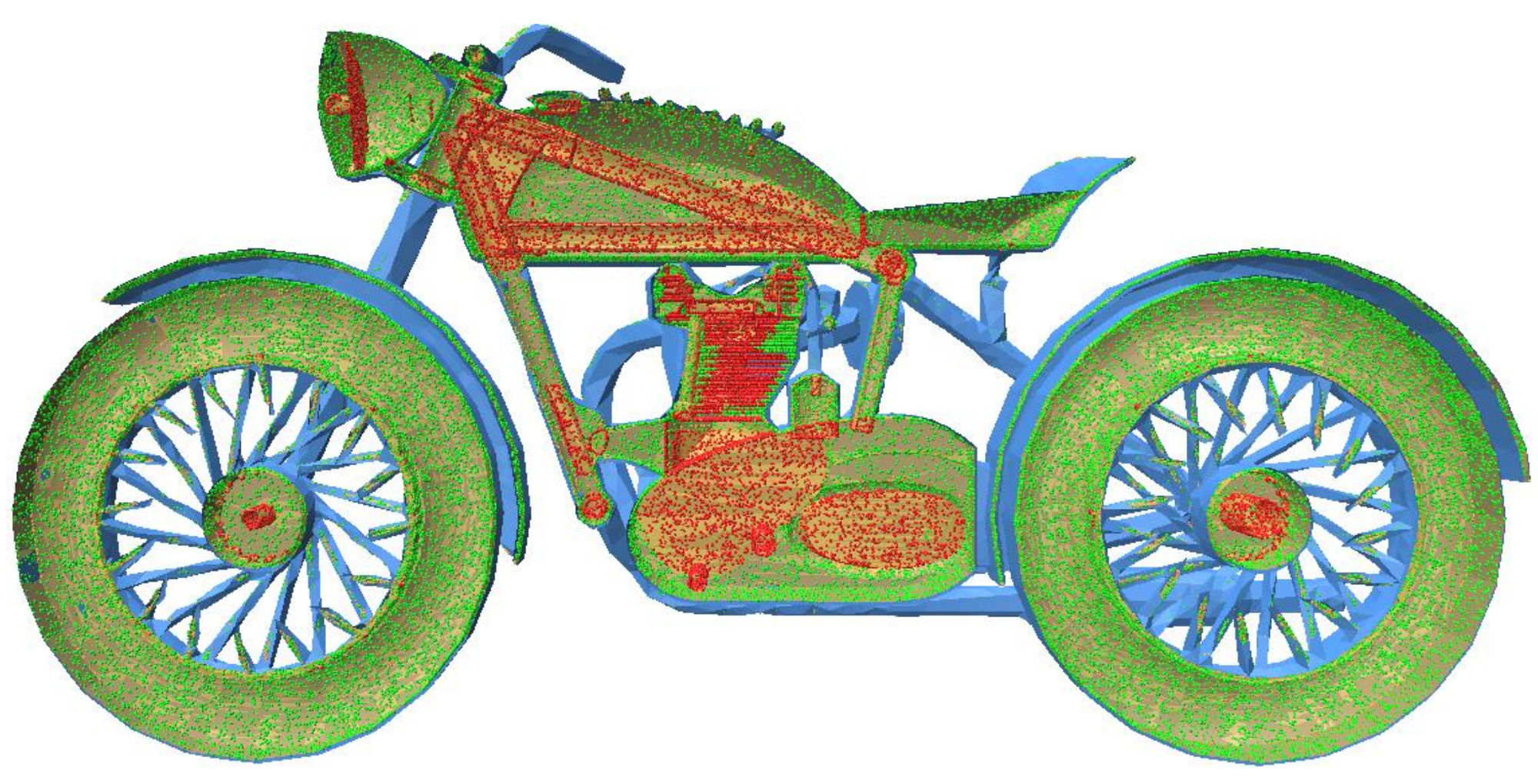}
    \end{subfigure}
    \caption{Extracting an oriented point cloud from the target mesh. Left: To clean the point cloud sampled on the preprocessed target mesh (orange), and shoot a beam of $N$ rays towards the initial mesh (blue), if more than half of the rays hit the initial mesh we keep the point (green) otherwise we prune it (red). Right: cross-section view of a Shapenet target mesh (orange, covered in samples) and the initial mesh (blue). Red points were flagged as inside and are omitted from the downstream steps.}
    \label{fig:prune_explain}
\end{figure}

Our planar projection loss (\Cref{sec:planar_projection}) utilizes a pointcloud-to-mesh projection scheme which relies on sampling the target mesh. These samples include both the point location and its parent face normal, which means ill-oriented faces pose a problem to this projection. To fix this, we start from the preprocessed mesh created for rendering (\Cref{sec:preprocess_rendering}), and pass it through an orientation procedure resembling \cite{cleanmesh}: we render the mesh from 36 different view points evenly spaced on a sphere, and count the number of pixels each face was visible from. We remove a face from the mesh if it doesn't meet a certain count threshold (see supplementary for further details), but not if the count is zero. Note that this means we don't keep obviously non-visible faces, but do preserve very small faces and internal structures. Then, we sample points uniformly at random on this mesh. To further clean the sampled point cloud, we use the initial mesh (\Cref{sec:init_mesh}). our insight here is that MANIFOLD outputs adhere to the general outline of the shape but are "inflated", we can use them to decide if a sampled point should be pruned or not. 
We first project each sample point $i$ to the initial mesh $p(i)$, and denote the projection vector length as $l_i$, the sampled points' parent face normals as $n_i$ and the projected points normal as $n_{p(i)}$. we then perform the following procedure:
We raycast a beam of $N$ rays (see supplementary for hyper-parameters settings used) from each sampled point in the direction of its projected point and check for intersections with the target mesh. If more than half the rays cast from a sampled point $i$ have a length larger than $l_i$, then we keep the sampled point (\Cref{fig:prune_explain}).

Additionally, to keep orientation consistent, we flip the sampled point normal for each point $i$ for which $n_i\cdot n_{p(i)} < 0.1$.

Finally, we project points from the initial mesh to the target mesh, in order to better sharp edges on the target mesh, see the supplementary for full details.

\section{Results}
\begin{figure}[ht]
        \includegraphics[width=0.49\textwidth]{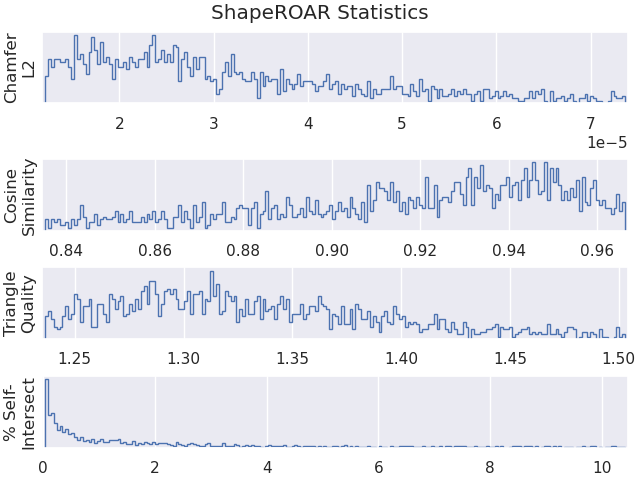}
        \caption{Statistical breakdown of 10k meshes from ShapeROAR, the reconstructed version of ShapeNet. Plots show histograms of several metrics. For a full description of the metrics see supplementary.}
        \label{fig:statistics}
\end{figure}
    
\begin{figure*}[ht]
\setlength{\tabcolsep}{0pt}
\begin{tabular}{cccccc}
Target & \begin{tabular}[c]{@{}c@{}}Continous\\Remeshing \cite{palfinger2022continuous} \end{tabular}   & Manifold+ \cite{manifold_plus} & fTet-Wild \cite{ftetwild} & \begin{tabular}[c]{@{}c@{}} Screened \\ Poisson \cite{screened_poisson} \end{tabular} & Ours \\[6pt]
\vspace{-10pt}
 \includegraphics[width=0.15\textwidth]{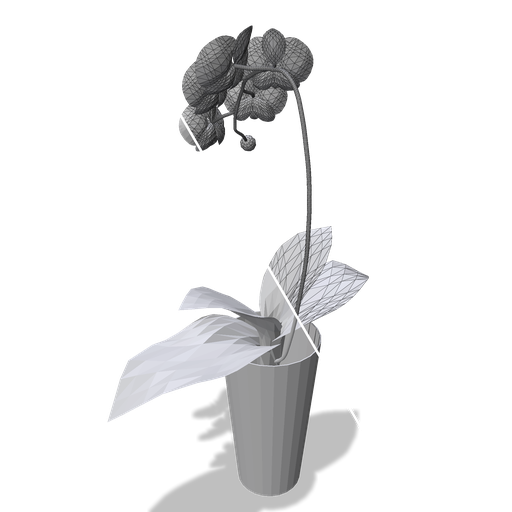} & \includegraphics[width=0.15\textwidth]{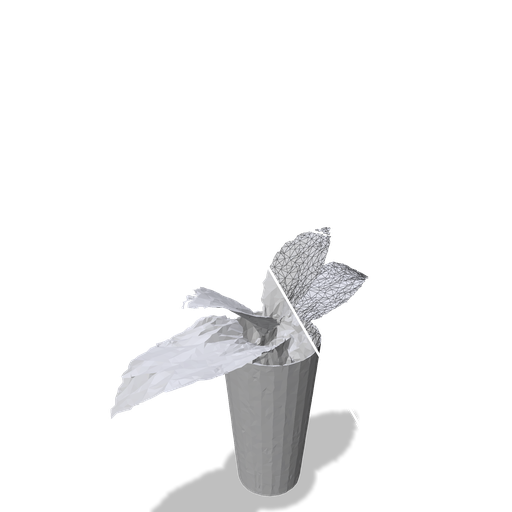} & \includegraphics[width=0.15\textwidth]{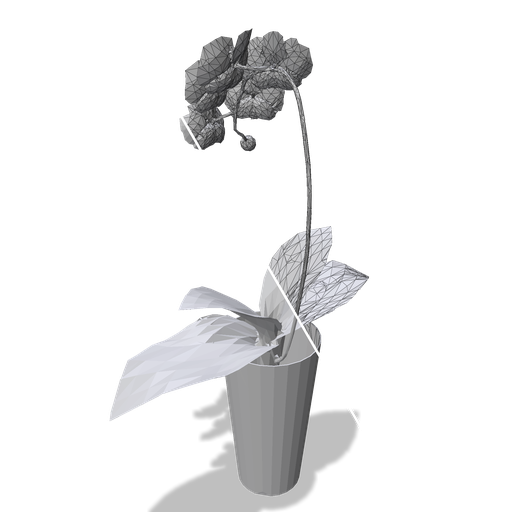} & \includegraphics[width=0.15\textwidth]{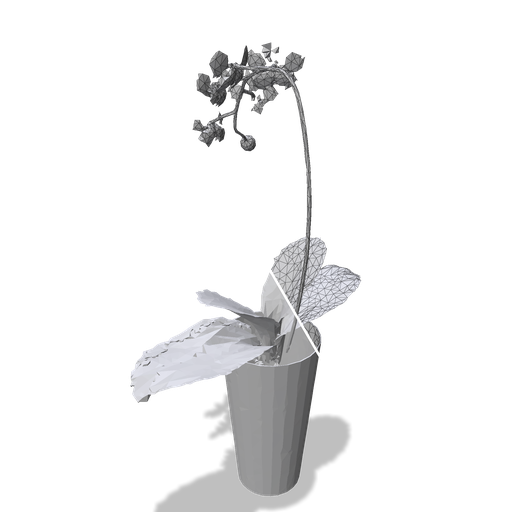} & \includegraphics[width=0.15\textwidth]{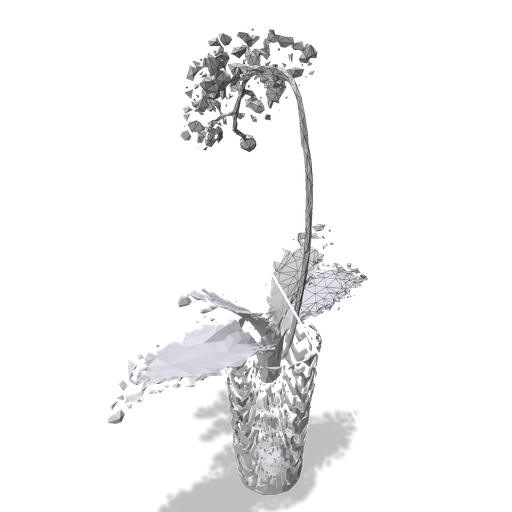} & \includegraphics[width=0.15\textwidth]{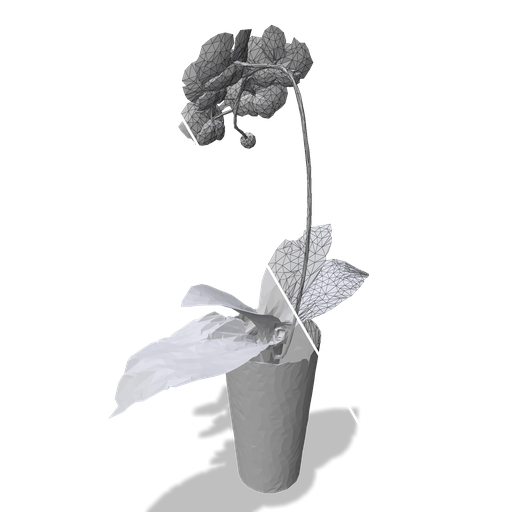} \\
\vspace{-30pt}
 \includegraphics[width=0.15\textwidth]{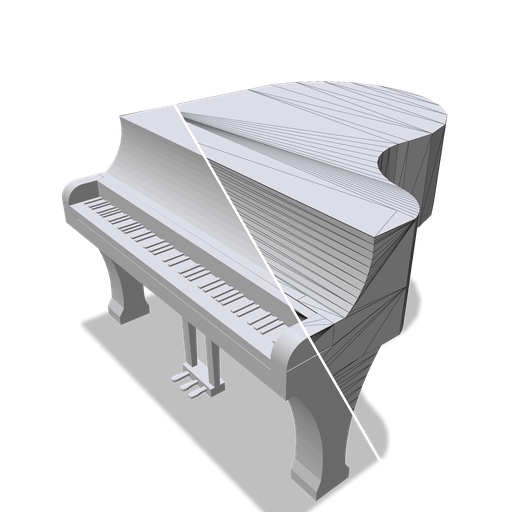} & \includegraphics[width=0.15\textwidth]{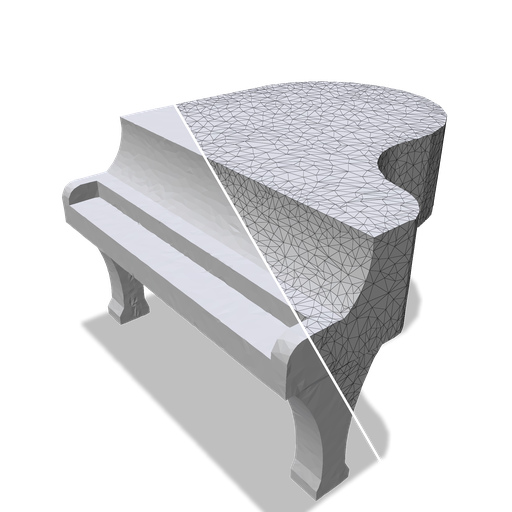} & \includegraphics[width=0.15\textwidth]{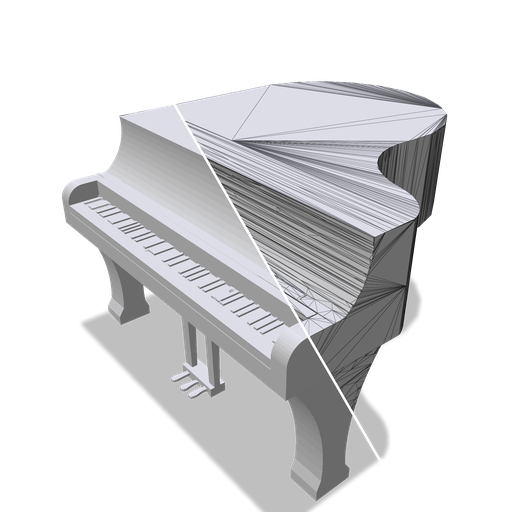} & \includegraphics[width=0.15\textwidth]{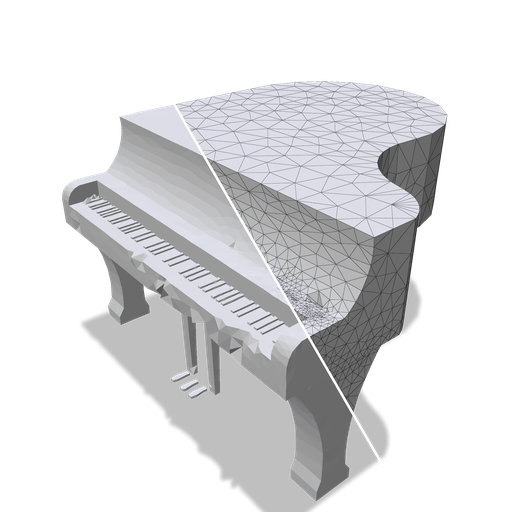} & \includegraphics[width=0.15\textwidth]{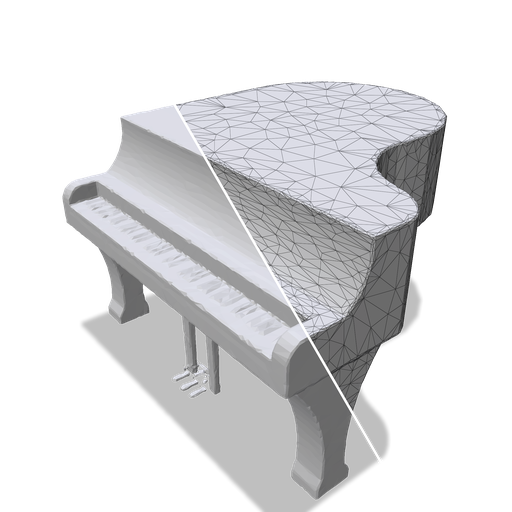} & \includegraphics[width=0.15\textwidth]{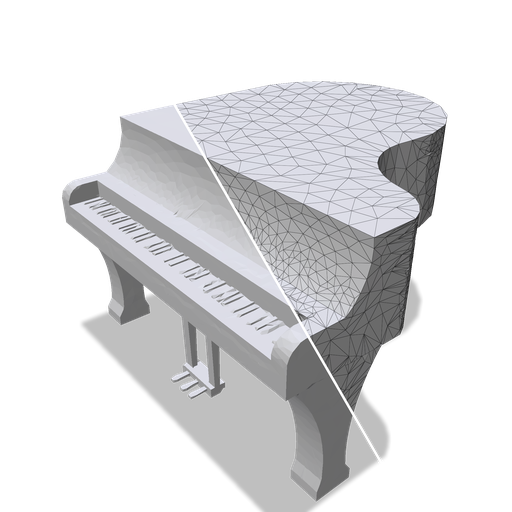} 
 \\
 \begin{tikzpicture}[spy using outlines={circle,yellow,magnification=3,size=1.5cm, connect spies}]
\node {\includegraphics[width=0.15\textwidth]{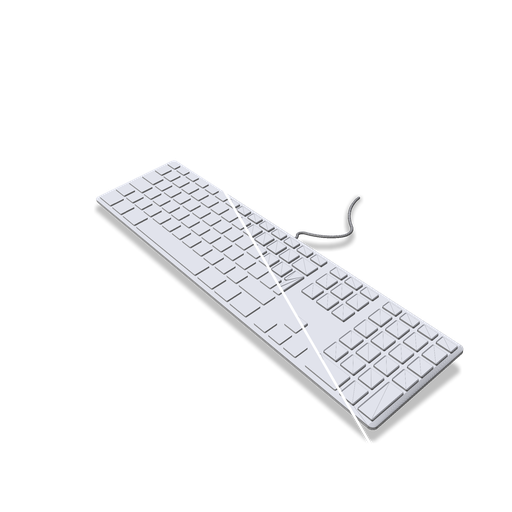}};
\spy on (-0.2,0.0) in node [left] at (1.0,-1.25);
\end{tikzpicture} & 
\begin{tikzpicture}[spy using outlines={circle,yellow,magnification=3,size=1.5cm, connect spies}]
\node {\includegraphics[width=0.15\textwidth]{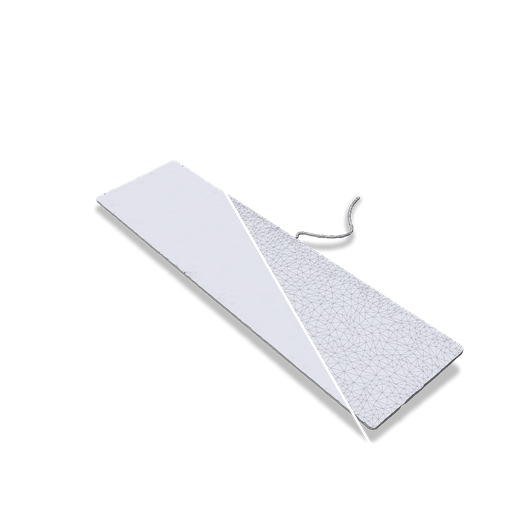}};
\spy on (-0.2,0.0) in node [left] at (1.0,-1.25);
\end{tikzpicture} &
\begin{tikzpicture}[spy using outlines={circle,yellow,magnification=3,size=1.5cm, connect spies}]
\node {\includegraphics[width=0.15\textwidth]{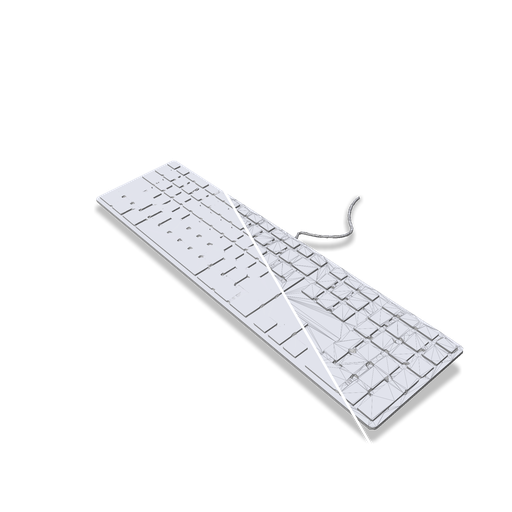}};
\spy on (-0.2,0.0) in node [left] at (1.0,-1.25);
\end{tikzpicture} &
\begin{tikzpicture}[spy using outlines={circle,yellow,magnification=3,size=1.5cm, connect spies}]
\node {\includegraphics[width=0.15\textwidth]{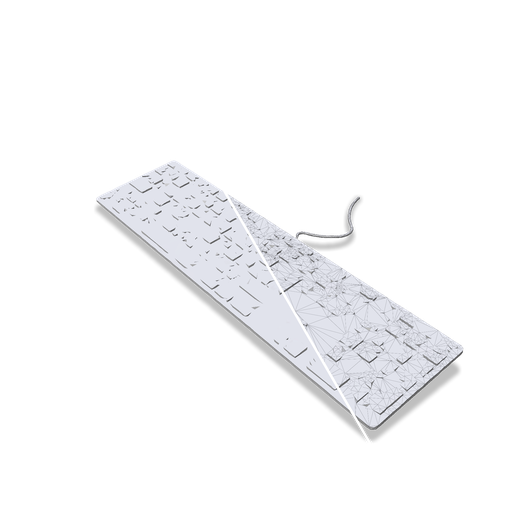}};
\spy on (-0.2,0.0) in node [left] at (1.0,-1.25);
\end{tikzpicture} &
\begin{tikzpicture}[spy using outlines={circle,yellow,magnification=3,size=1.5cm, connect spies}]
\node {\includegraphics[width=0.15\textwidth]{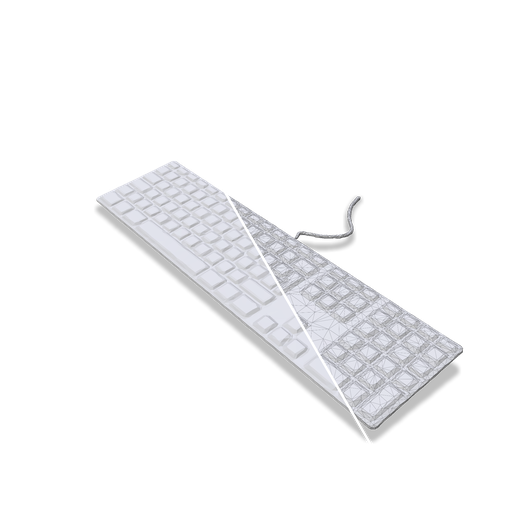}};
\spy on (-0.2,0.0) in node [left] at (1.0,-1.25);
\end{tikzpicture} &
\begin{tikzpicture}[spy using outlines={circle,yellow,magnification=3,size=1.5cm, connect spies}]
\node {\includegraphics[width=0.15\textwidth]{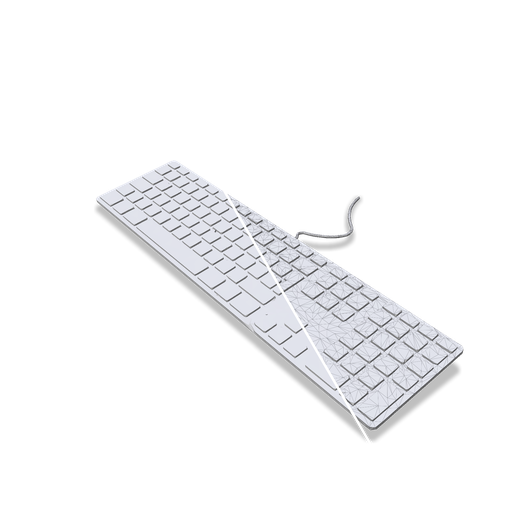}};
\spy on (-0.2,0.0) in node [left] at (1.0,-1.25);
\end{tikzpicture} \\
\end{tabular}
\caption{Qualitative comparisons of our method versus previous methods for reconstructing objects from the ShapeNet dataset. Our method preserves geometry better than competitive methods, with much better triangle quality, and a 2-manifold output.}
\label{fig:shapnet_qual}
\end{figure*}
\subsection{Reconstruction in-the-wild}
\label{sec:in-the-wild}
    In order to show our methods ability to reconstruct realistic 3D mesh data, we experiment with reconstructing the ShapeNet dataset \cite{shapenet2015}. A statistical breakdown of some metrics can be seen in \Cref{fig:statistics}. Additionally, a randomly selected subset of 10 meshes per class was used for quantitative comparisons (as many of the other methods were too slow to process a larger portion of the data). Results can be seen in \Cref{tab:shapenet_all} and \Cref{fig:shapnet_qual}. 
    The triangle meshes in the ShapeNet subset are particularly hard to process due to their low quality: despite the seemingly coherent appearance of the models, they suffer from acute connectivity problems, including disconnected components, self-intersections, inconsistent normals, duplicated and degenerate faces, un-referenced vertices and highly irregular vertex neighborhoods. To deal with these issues, we pass all the meshes through a cleaning pre-processing step described in \Cref{sec:orient_target}. We allow other methods to benefit from this cleaning scheme as well if it was observed to improve their results, and for point-cloud based methods we input the cleaned and oriented point cloud produced by our method. For fairness, the same point cloud is used for our projection step in the Face Split block and the planar projection loss (\Cref{sec:planar_projection}).The initial source mesh used in our method for optimization is created using the procedure described in \Cref{sec:init_mesh}. We observed better results when setting our triangle budget to a high value (50k faces) and decimating it \cite{qslim, meshlab} as a post process step to meet a certain budget (see supplementary for additional detail). Other methods were also decimated using the same method if it proved beneficial for their results rather than setting a target budget. Some methods were allowed to run with an unconstrained triangle budget if they do not support setting a budget directly and decimation was observed to significantly harm their performance. We found the volumetric based methods to be competitive, but fail to deliver on their promise for 2-manifold results. Other methods do not capture the surface geometry as well or take large amount of time ($\sim 10$ hours) to process a mesh \cite{MeshRepair}.
    
    \textbf{Range Scan Reconstruction} In addition, we have experimented with range scans from the Stanford Scan Repository \cite{armadillo, bunny} (see supplementary for qualitative results). These scans are typically quite noisy, and lack regions where the scanning operation failed, leaving holes in the scans resulting geometry. Our preliminary result showed promising results, reconstructing the shape reasonably well. This suggests our method is quite robust to rendering artifacts.. 

\begin{table*}
\caption{Performance comparisons over 550 meshes (10 per class) in ShapenetCoreV2.}
\footnotesize
\begin{tabularx}{\textwidth}{X|X|X|X|X|X|X}
\hline
Method & Chamfer L2 \newline ($\cdot1e^{-4}$) $\downarrow$ & Cosine Similarity ($\cdot1e^{-1}$) $\uparrow$ & Triangle Quality $\downarrow$ & \% Self Intersect $\downarrow$ & \% Non Manifold Edges $\downarrow$ & \% Non Manifold Vertices ($\cdot1e^{-2}$) $\downarrow$ \\
\hline\hline
\multicolumn{7}{c}{Face Budget 500/1k/5k/10k} \\
\hline\hline
Ours & \textbf{2.9}/\textbf{1.5}/\textbf{0.51}/\textbf{0.44} & 8.6/8.8/9.0/9.1 & \textbf{1.9}/\textbf{1.8}/\textbf{1.5}/\textbf{1.4} & 10/7.0/4.0/3.9 & \textbf{0} & \textbf{0}\\
\hline
Man. \cite{manifold} & 6.3/5.6/5.0/5.0 & \textbf{8.9}/\textbf{9.0}/9.1/9.2 & 2.7/2.5/1.7/1.5 & \textbf{2.6}/\textbf{1.8}/\textbf{0.59}/\textbf{0.43} & \textbf{0} & 5.6/13/3.5/1.6\\
\hline
Man.+ \cite{manifold_plus} & 61/19/2/1.8 & 8.4/8.9/\textbf{9.3}/\textbf{9.3} & 494/45/7.0/5.6 & 8.1/6.3/4.7/4.3 & 3.3/1.8/0.46/0.26 & 37/32/16/14\\
\hline\hline
\multicolumn{7}{c}{Face Budget $\geq$ 10k} \\
\hline\hline
Cont. Re \cite{palfinger2022continuous} & 49 & 8.8 & 1.0 & 1.9 & 0.0017 & 0.044\\
\hline
TetW \cite{tetwild} & 107 & 8.8 & 1.2 & 0.0034 & 1.9 & 42\\
\hline
fTetW \cite{ftetwild} & 88 & 8.9 & 1.2 & 0.0018 & 2.3 & 48\\
\hline
RIMLS \cite{RIMLS_marching} & 13 & 9.4 & 1.2 & 4.8 & 3.0 & 19\\
\hline
P2M \cite{point2mesh} & 2208 & 2.9 & 1.6 & 25 & 0 & 0.020\\
\hline
S. Poisson \cite{screened_poisson} & 8.5 & 8.9 & 1.2 & 3.7$\cdot1e^{-5}$ & 0 & 0\\
\hline
3D A. Wrap \cite{alphawrap_cgal} & 20 & 9.3 & 1.1 & 0.036 & 0 & 0\\

\end{tabularx}

\label{tab:shapenet_all}
\end{table*} %

\subsection{Neural SDF Reconstruction}
\begin{figure}[ht]
    \includegraphics[width=0.49\textwidth]{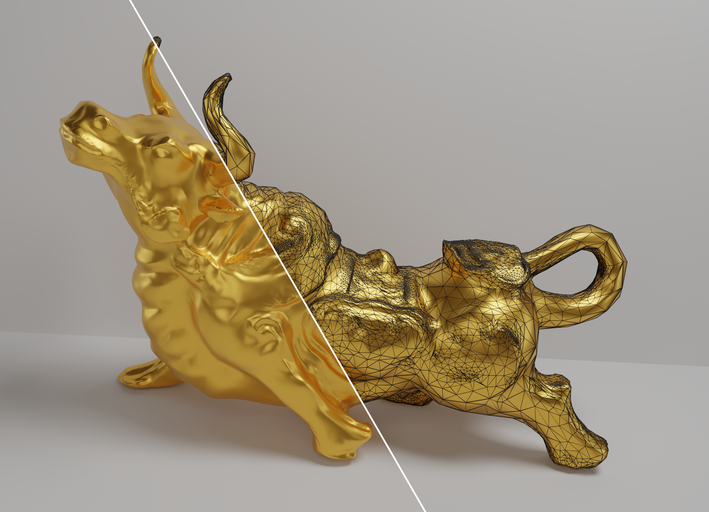}
    \caption{ROAR reconstruction of \textbf{Bull} from a Neural SDF. Our method can be applied to any 3D shape that is renderable and for which a projection operation can be defined.}
    \label{fig:bull}
\end{figure}

    Since our method relies on a render loss and a valid projection operation, it can be easily extended to work with signed distance functions, as their zero level set can also be rendered with normals encoded as color, and projection of a point to a level set consists of computing the SDF and normal: $v_{proj} = v -\frac{\nabla{SDF(v)}}{||\nabla{SDF(v)}||} \cdot SDF(v)$. We used a set of volSDF \cite{volSDF} models, trained over a subset of the BlendedMVS\cite{blendedMVS} dataset and compared our method's results to using a (double) marching cubes approach as implemented by the original authors for a given triangle budget. Results can be seen in \Cref{tab:SDF} and \Cref{fig:bull}. The double MC approach entails performing a coarse MC reconstruction to eliminate disconnected components and to shrink the bounding box around the object of interest, followed by a fine resolution MC reconstruction yielding the final result. It was decimated to meet the triangle budget using Qslim \cite{qslim}. We used the same camera viewpoint parameters as provided in the dataset. To render our target from different views, we implemented a simple sphere tracer and used its renders as the target images for the $L_{Im}$ loss term in \Cref{eq:loss}. The planar projection term $L_{Proj}$ was replaced with the mean distance between all vertices and their projection to the zero level set. Notice that while our results are on par in terms of Chamfer distance, our method performs roughly an order of magnitude less SDF evaluations for computation (that are due to rendering the SDF, and projection) with no optimizations, compared to the $512^3$ grid used for MC. In other words, our method is more efficient in extracting roughly the same information. Note that the Robot scene had a large piece of its wing missing in the ground truth mesh. The neural SDF did capture it and we reconstructed the wing fully (yet got penalized for it), as opposed to marching cubes which failed to reconstruct it and achieved a higher score.

\begin{table*}[ht]
    \caption{Performance comparisons for the reconstruction of 9 neural SDFs from BlendedMVS for a 50k face budget.}
    \setlength\tabcolsep{6pt} 
        \aboverulesep=0ex
        \belowrulesep=0ex
        \renewcommand{\arraystretch}{1.5}
        \begin{tabular}[t]{@{\hskip0.0pt}llccccccccc@{\hskip0.0pt}}
            \toprule
            & Scene & Doll & Egg & Head & Angel & Bull & Robot & Dog & Bread & Camera\\
         \toprule
         \multirow{2}{*}{Chamfer $L_2$}
         & Ours &2.2e-5 & \textbf{1.9e-5} & 5.0e-6 & \textbf{0.76e-3} & \textbf{1.6e-3} & 0.016 & 3.1e-4 & \textbf{0.98e-5} & \textbf{8.3e-2}  \\
         & VolSDF\cite{volSDF} &  \textbf{1.6e-5} & 2.2e-5 & \textbf{3.5e-6} & 2.3e-3 & 1.7e-3 & \textbf{0.0018} & \textbf{1.5e-4} & 1.8e-5 & 8.7e-2  \\
        
            
         \midrule
         \multirow{2}{*}{SDF evals}
         & Ours & \textbf{46M} & \textbf{40M} & \textbf{27M} & \textbf{55M} & \textbf{41M} & \textbf{70M} & \textbf{18M} & \textbf{32M} & \textbf{96M}  \\
         & VolSDF\cite{volSDF} & \multicolumn{9}{c}{270M}  \\
         
        \bottomrule
        \end{tabular}
        \vspace{3pt}   
    \label{tab:SDF}
\end{table*}

\subsection{Ablations}
    \label{sec:ablations}
\begin{figure*}
    \centering
    \begin{subfigure}{0.15\textwidth}
    \begin{tikzpicture}[spy using outlines={circle,yellow,magnification=3,size=1.5cm, connect spies}]
    \node {\includegraphics[width=\textwidth]{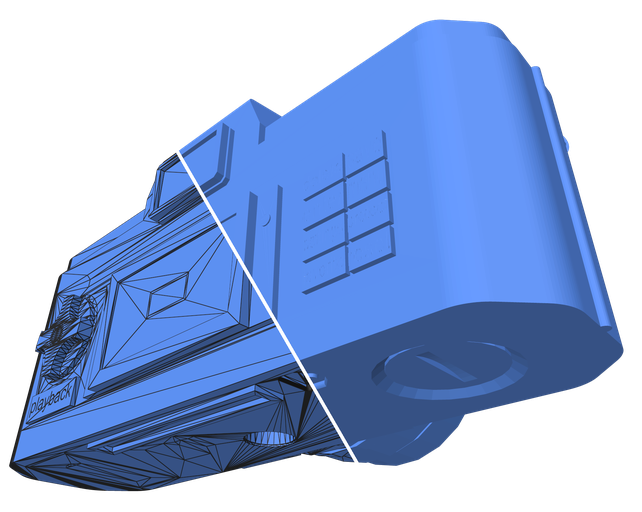}};
    \spy on (0.1,0.1) in node [left] at (1.0,-1.25);
    \end{tikzpicture}
    
    \caption{Target}
    \label{fig:proj1}
    \end{subfigure}
    \begin{subfigure}{0.15\textwidth}
    \begin{tikzpicture}[spy using outlines={circle,yellow,magnification=3,size=1.5cm, connect spies}]
    \node {\includegraphics[width=\textwidth]{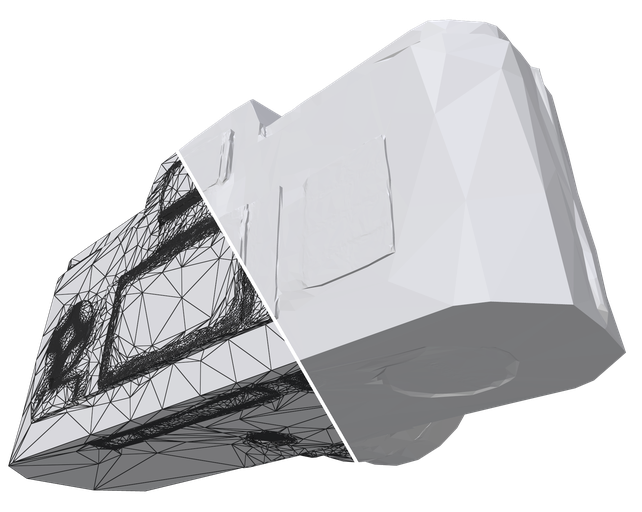}};
    \spy on (0.1,0.1) in node [left] at (1.0,-1.25);
    \end{tikzpicture}
    
    \caption{No P.P. Loss}
    \label{fig:proj2}
    \end{subfigure}
    \begin{subfigure}{0.15\textwidth}
    \begin{tikzpicture}[spy using outlines={circle,yellow,magnification=3,size=1.5cm, connect spies}]
    \node {\includegraphics[width=\textwidth]{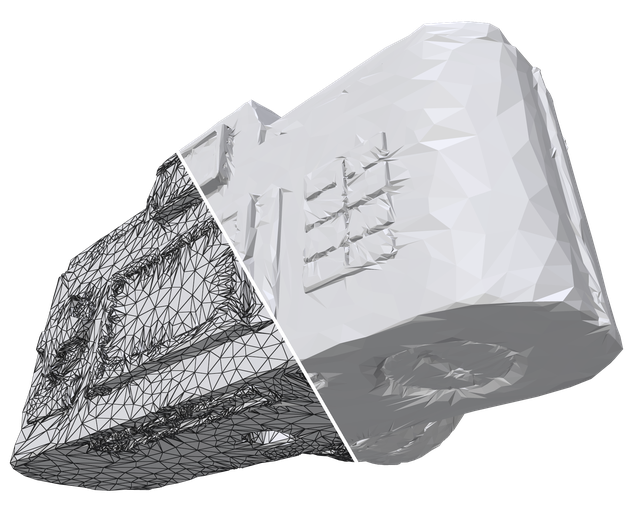}};
    \spy on (0.1,0.1) in node [left] at (1.0,-1.25);
    \end{tikzpicture}
    
    \caption{No Render Loss}
    \label{fig:proj3}
    \end{subfigure}
    \begin{subfigure}{0.15\textwidth}
    \begin{tikzpicture}[spy using outlines={circle,yellow,magnification=3,size=1.5cm, connect spies}]
    \node {\includegraphics[width=\textwidth]{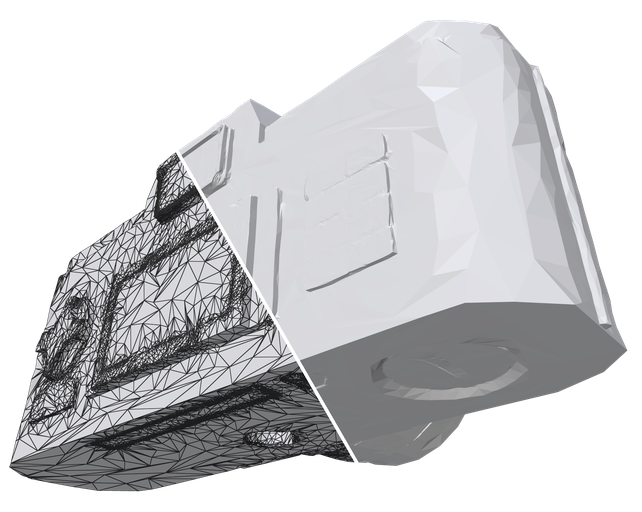}};
    \spy on (0.1,0.1) in node [left] at (1.0,-1.25);
    \end{tikzpicture}
    
    \caption{Chamfer}
    \label{fig:proj4}
    \end{subfigure}
    \begin{subfigure}{0.15\textwidth}
    \begin{tikzpicture}[spy using outlines={circle,yellow,magnification=3,size=1.5cm, connect spies}]
    \node {\includegraphics[width=\textwidth]{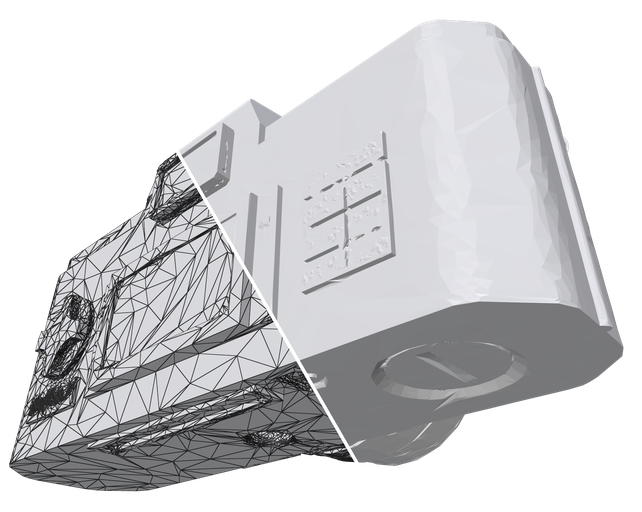}};
    \spy on (0.1,0.1) in node [left] at (1.0,-1.25);
    \end{tikzpicture}
    
    \caption{Ours, 1NN}
    \label{fig:proj5}
    \end{subfigure}
    \begin{subfigure}{0.15\textwidth}
    \begin{tikzpicture}[spy using outlines={circle,yellow,magnification=3,size=1.5cm, connect spies}]
    \node {\includegraphics[width=\textwidth]{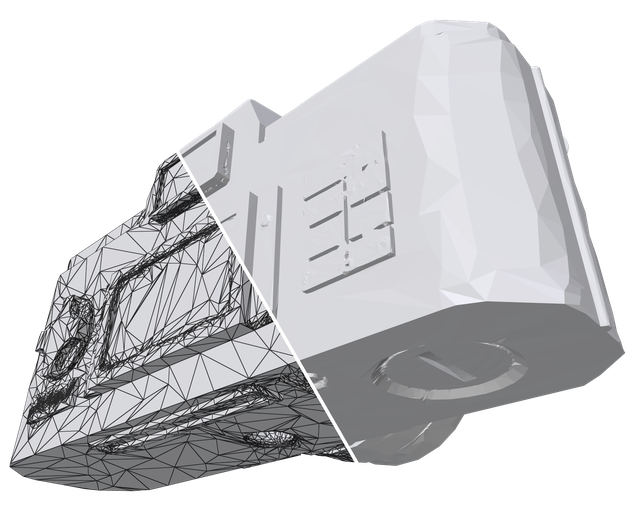}};
    \spy on (0.1,0.1) in node [left] at (1.0,-1.25);
    \end{tikzpicture}
    
    \caption{Ours, 5NN}
    \label{fig:proj6}
    \end{subfigure}
    \caption{Loss Term Ablation. We ran our method with different loss configurations. (a) The target mesh. (b) Without planar projection. (c) Without render loss. (d) With Chamfer loss replacing planar projection. (e) With planar projection loss, set to average over 1 NN. (f) Our default setting of 5NN.}
    \label{fig:projection_loss}
\end{figure*}

    A quantitative ablation over 11 high resolution water tight meshes was conducted (See \Cref{tab:ablations}), where the target face budget was set to 10k. We chose these meshes because they allow for better quantitative evaluation of shape reconstruction rather than using a triangle soup as input where the underlying shape is ill-defined. Because the meshes were clean and of a relatively organic nature (see supplementary), the planar projection loss term was not used for optimization and its effects were tested separately (\Cref{fig:projection_loss}). We froze all hyper parameters and computed the image loss (16 novel view points) and Chamfer distance (200k samples) from the ground truth in the following settings: Full - our full pipeline engaged, Silhouette - the rendering is performed with binary silhouettes only instead of normals encoded as colors, Half Views - we render only 18 views instead of 36, No T. Smooth - we do not perform tangential smoothing nor edge flips, No FC - we do not run the Face collapse block, No $l_{att}$ - we set $l_{att}=1$ effectively removing it, $l_{att}$ CLC - we set $l_{att}$ = $l_{ref}$ (\Cref{eq:vertex_update_step}), Max Dist - we replace the criterion in \Cref{eq:face_curvature} to instead using the maximum distance of super-samples to the target samples and use that as face scores which proved to be a much poorer estimator of local geometry changes.

    \subsubsection{GPU speedup}
\begin{figure}[h]
        \includegraphics[width=0.49\textwidth]{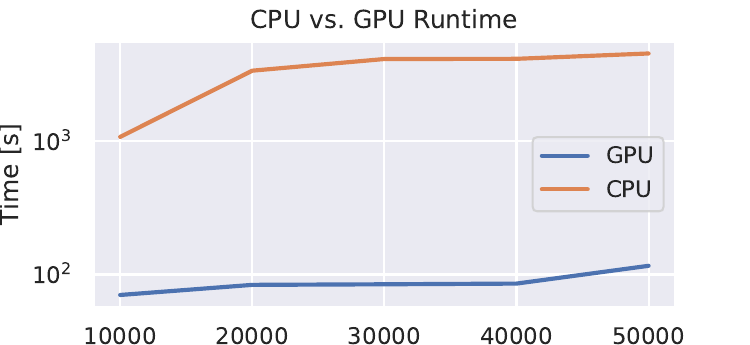}
        \caption{Average Run time for 5 random ShapeNet Meshes, for 5 different face budget levels, on both the CPU and the GPU. The CPU run time is about an order of magnitude ($\sim$hour) larger than the GPU run time ($\sim$minute)}.
        \label{fig:timings}
    \end{figure}

    To demonstrate that our system is optimized for GPU usage, we ran several examples both on the GPU (GeForce RTX 2080 Ti) and CPU (Intel(R) Xeon(R) Silver 4114 CPU @ 2.20GHz) and quantify the run time difference. We selected 5 random ShapeNet meshes from different classes, and run our pipeline for 5 different face budgets. The results is shown in \Cref{fig:timings}.
\begin{table}[ht]
    \caption{Ablation table of our method over 11 high resolution meshes with a face budget of 10k}
    \begin{tabularx}{0.5\textwidth}{X|X|X}
    \hline
    Method & Image L1 ($\cdot1e^{-2}$) $\downarrow$ & Chamfer $L_2$ ($\cdot1e^{-4}$) $\downarrow$ \\
    \hline\hline
    Full & 1.06 & 0.967 \\
    \hline
    Silhouette & 3.58 & 53.5 \\
    \hline
    Half Views \# views & 1.24 & 1.65 \\
    \hline
    No T. Smooth & 1.23 & 1.31 \\
    \hline
    No FC & 1.19 & 1.21 \\
    \hline
    No $l_{att}$ & 5.64 & 40.1 \\
    \hline
    $l_{att}$ CLC & 1.32 & 1.51 \\
    \hline
    Max Dist & 2.58 & 28.8 \\
    \end{tabularx}
    \bigskip\centering
    \label{tab:ablations}
\end{table}

\begin{figure*}
    \centering
    \begin{subfigure}{0.15\textwidth}
    \includegraphics[width=\textwidth]{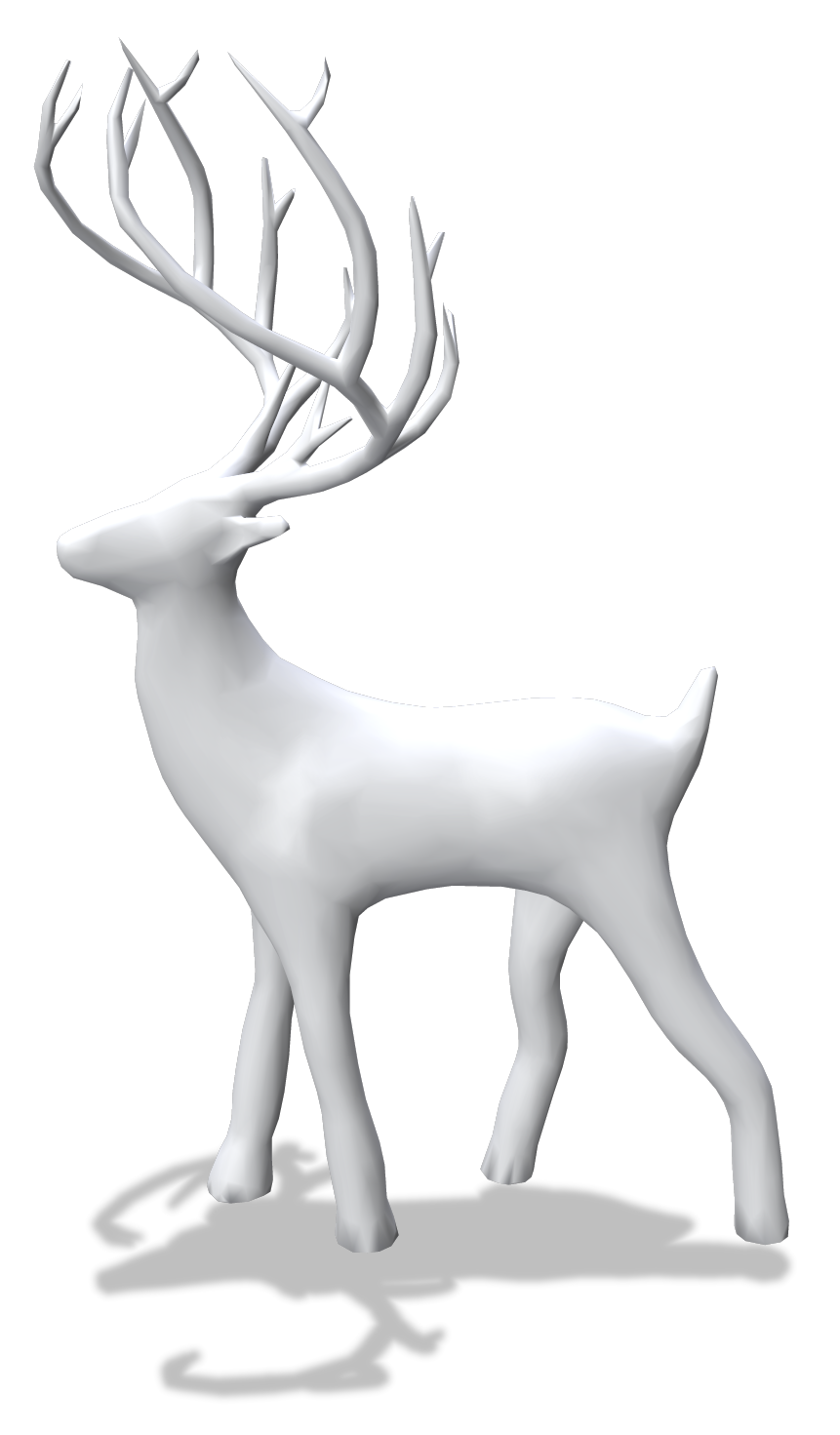}
    \caption{Target}
    \label{fig:ablation1}
    \end{subfigure}
    \begin{subfigure}{0.15\textwidth}
    \includegraphics[width=\textwidth]{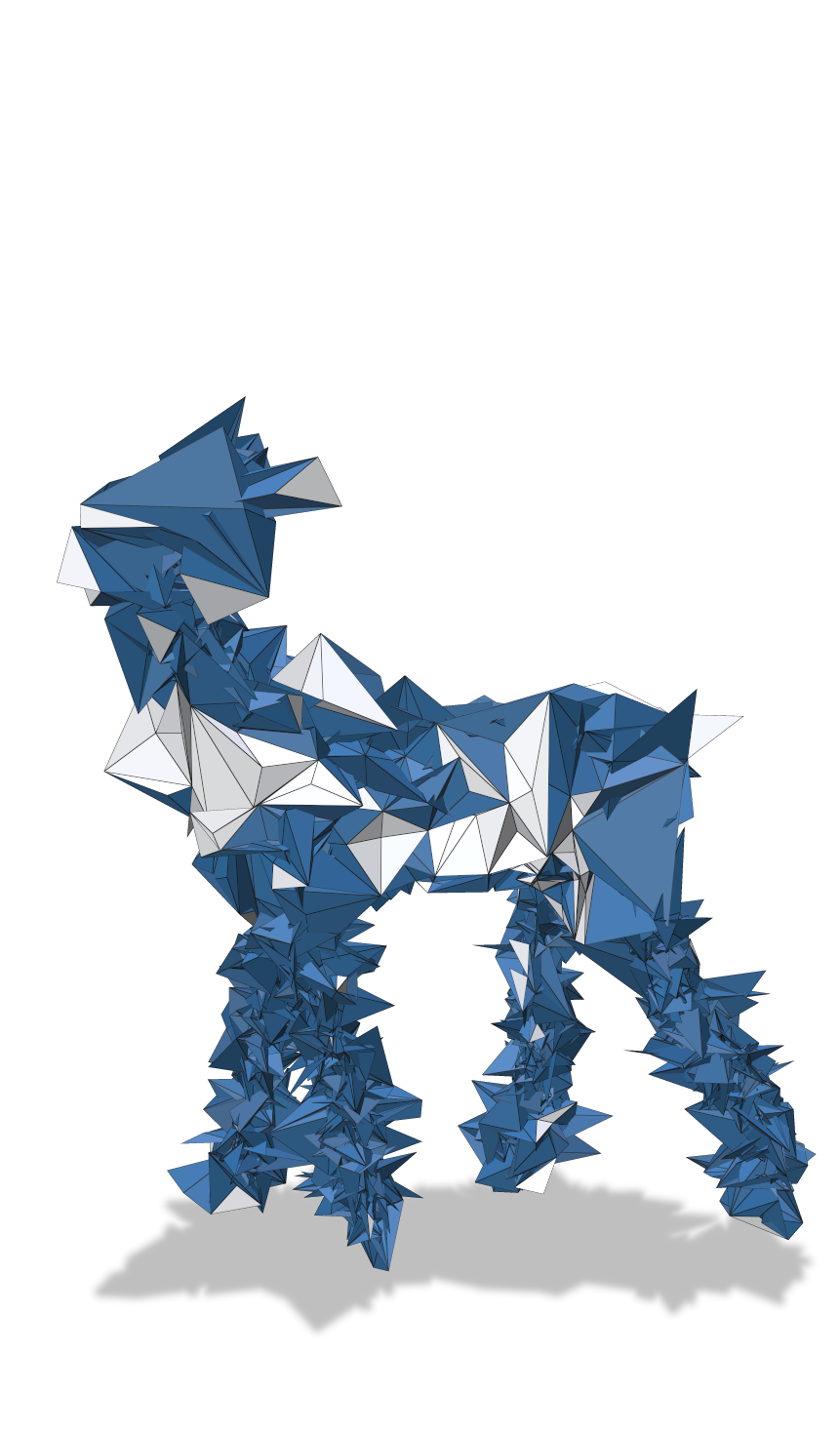}
    \caption{No $l_{att}$}
    \label{fig:ablation2}
    \end{subfigure}
    \begin{subfigure}{0.15\textwidth}
    \begin{tikzpicture}[spy using outlines={circle,yellow,magnification=3,size=1.5cm, connect spies}]
    \node {\includegraphics[width=\textwidth]{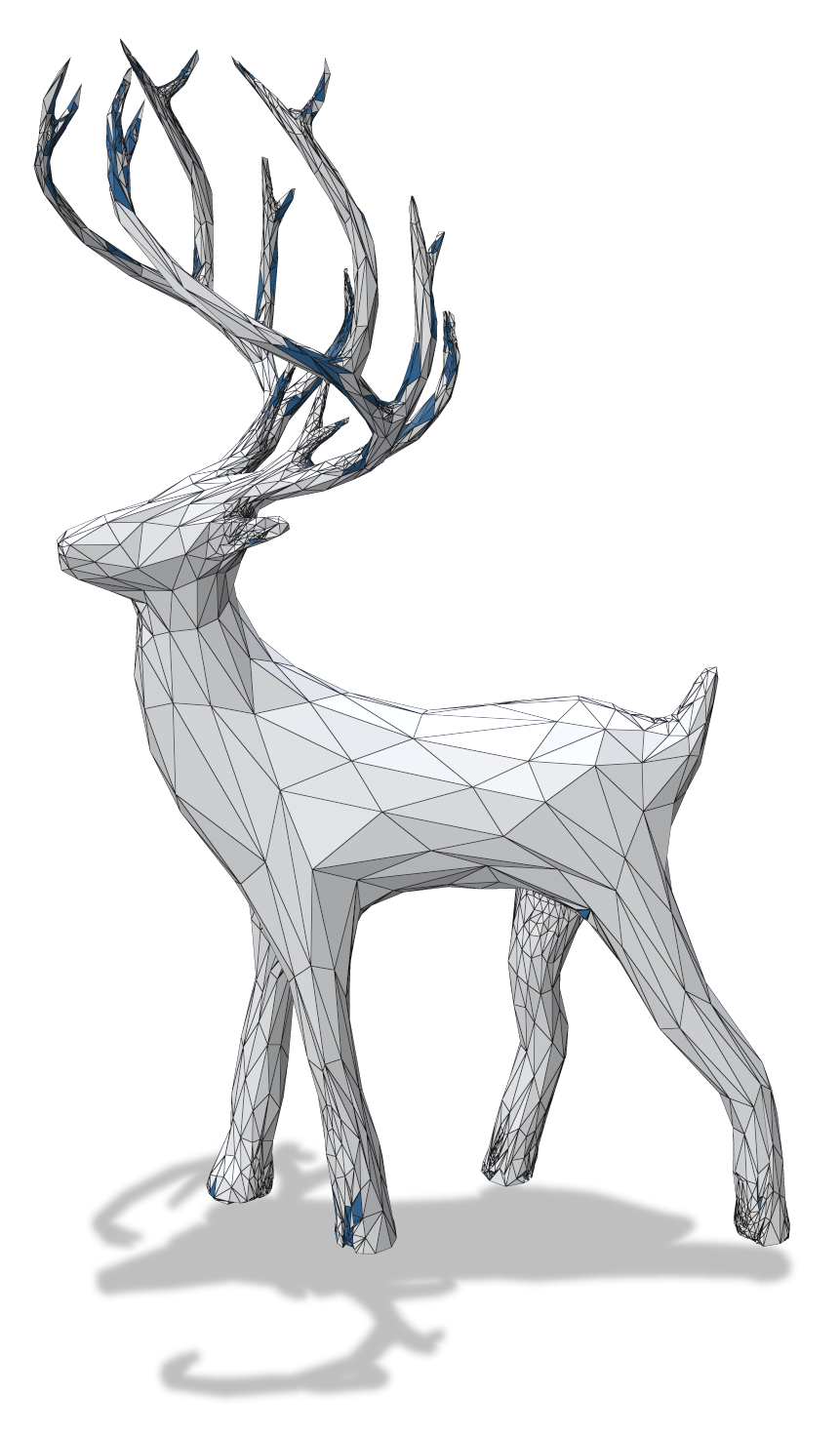}};
    \spy on (-0.3,1.2) in node [left] at (-0.2,-0.25);
    \end{tikzpicture}
    \caption{No Face Collapse}
    \label{fig:ablation4}
    \end{subfigure}
    \begin{subfigure}{0.15\textwidth}
    \includegraphics[width=\textwidth]{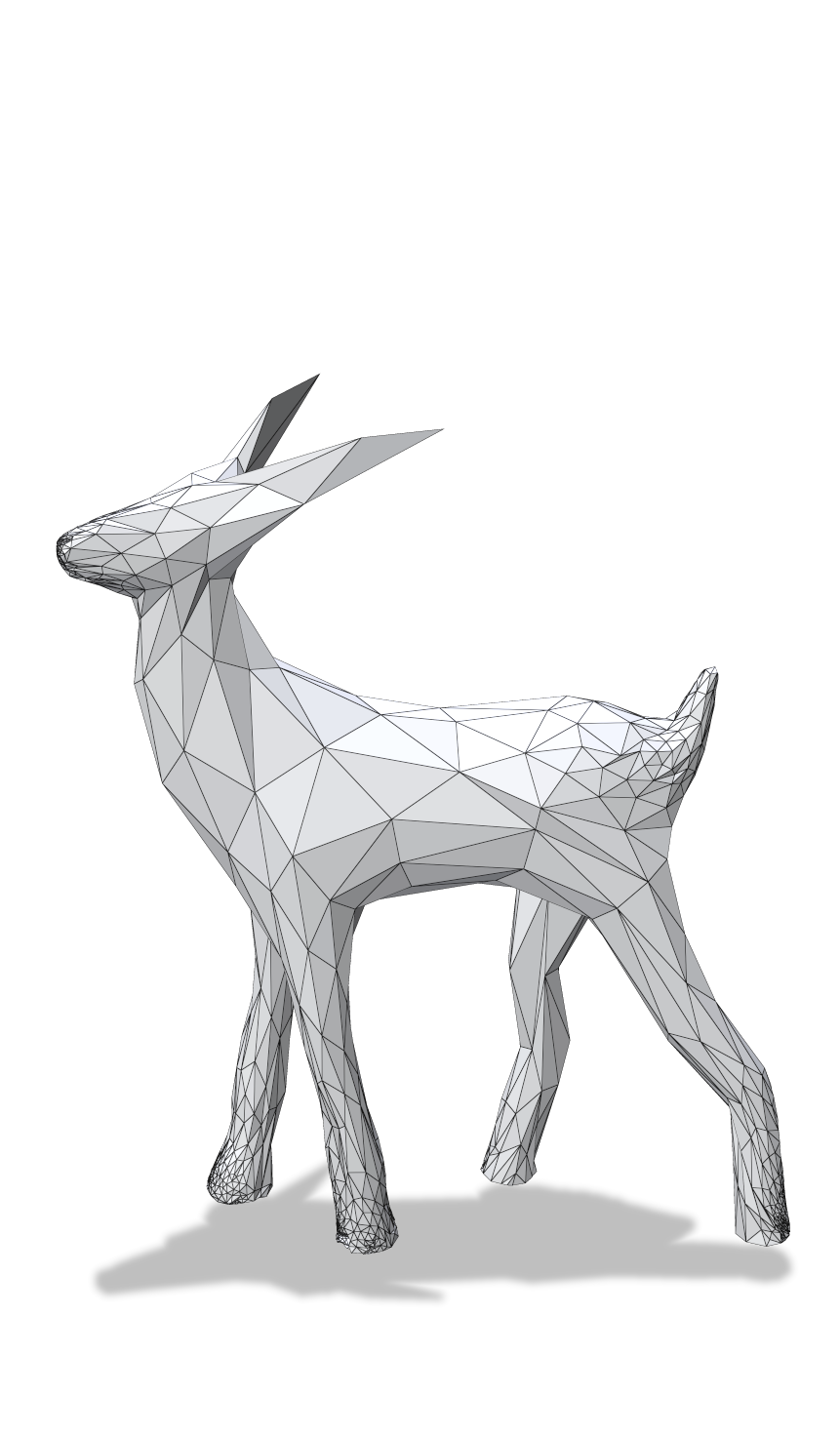}
    \caption{Max Dist.}
    \label{fig:ablation5}
    \end{subfigure}
    \begin{subfigure}{0.15\textwidth}
    \includegraphics[width=\textwidth]{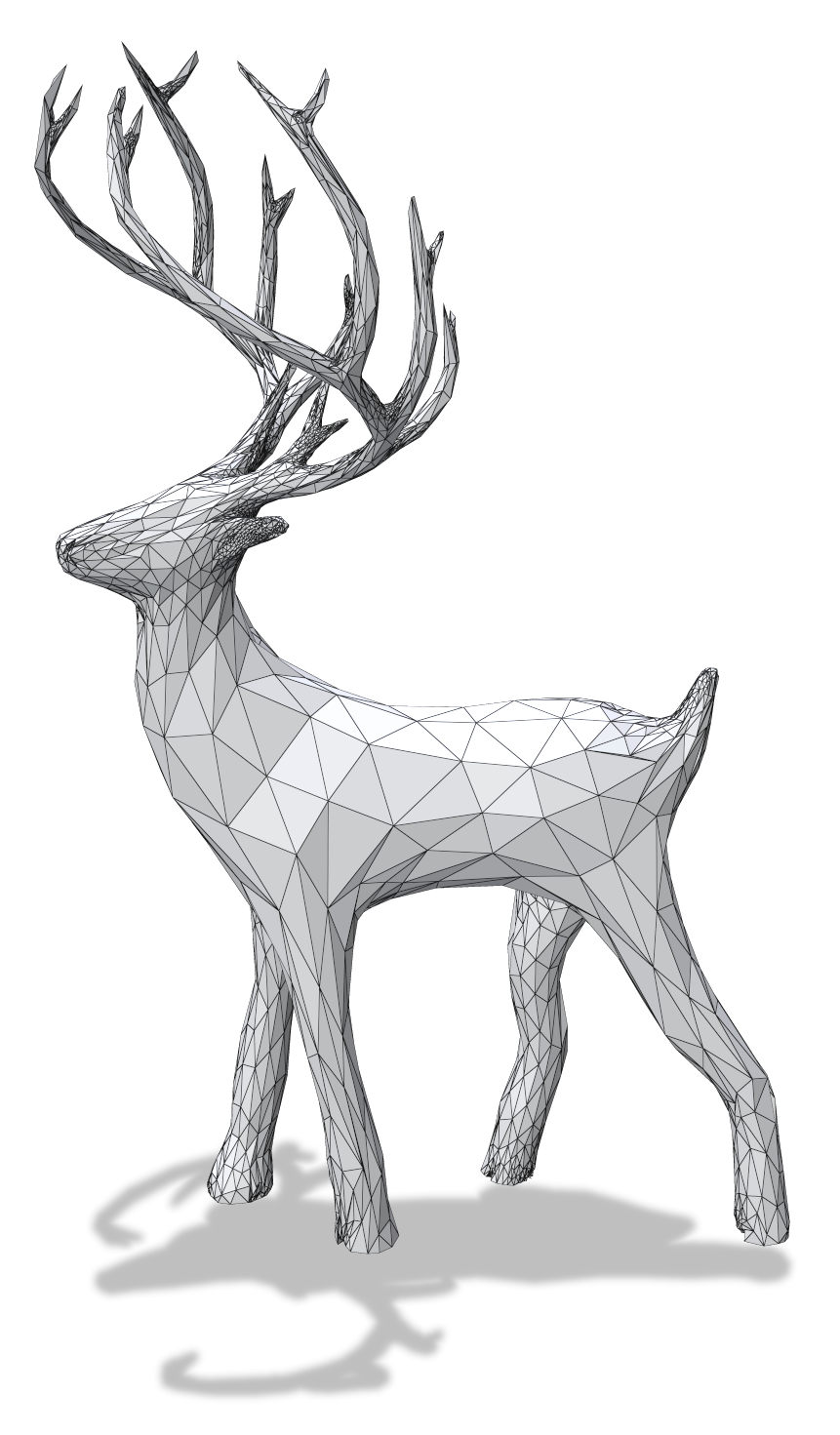}
    \caption{Ours}
    \label{fig:ablation6}
    \end{subfigure}
    \caption{Qualitative Block Ablation. We remove specific blocks from the pipeline and test the result (self intersections in blue). (a) Target mesh. (b) Setting $l_{att}$ to 1. (c) Removing the Face Collapse block (\Cref{sec:face_collapse}). (d) A more naive criterion for face curvature estimation is used: the maximum distance to the nearest neighbors. (e) Full pipeline.}
    \label{fig:ablations}
\end{figure*}

\section{Conclusions}
ROAR is the first approach to offer a full GPU-based mesh evolution process that is topologically error free in practice. The differentiable nature of our system makes it possible to be used in conjecture with advanced solvers such as ADAM. 

Our main result, ShapeROAR, is both important on its own for the field of geometric deep learning, and demonstrates the unprecedented robustness and feasibility of our approach. As shown by our experiments, our carefully designed steps fit together like puzzle pieces to yield the right balance between topological correctness, reconstruction quality, and triangulation quality.  

We believe this robust approach of mixed 2D and 3D considerations, and a mechanism to control and correct the triangulation during mesh evolution, is mature and stable enough to facilitate meaningful 3D learning.  In light of the largely available yet broken shape datasets, we anticipate a comeback for the triangular mesh to the forefront of geometric deep learning, fully unlocking the potential of this representation.  

\begin{figure}[ht]
\begin{tabular}{cc}
\begin{tikzpicture}[spy using outlines={circle,yellow,magnification=4,size=2.0cm, connect spies}]
    \node {\includegraphics[width=0.1\textwidth]{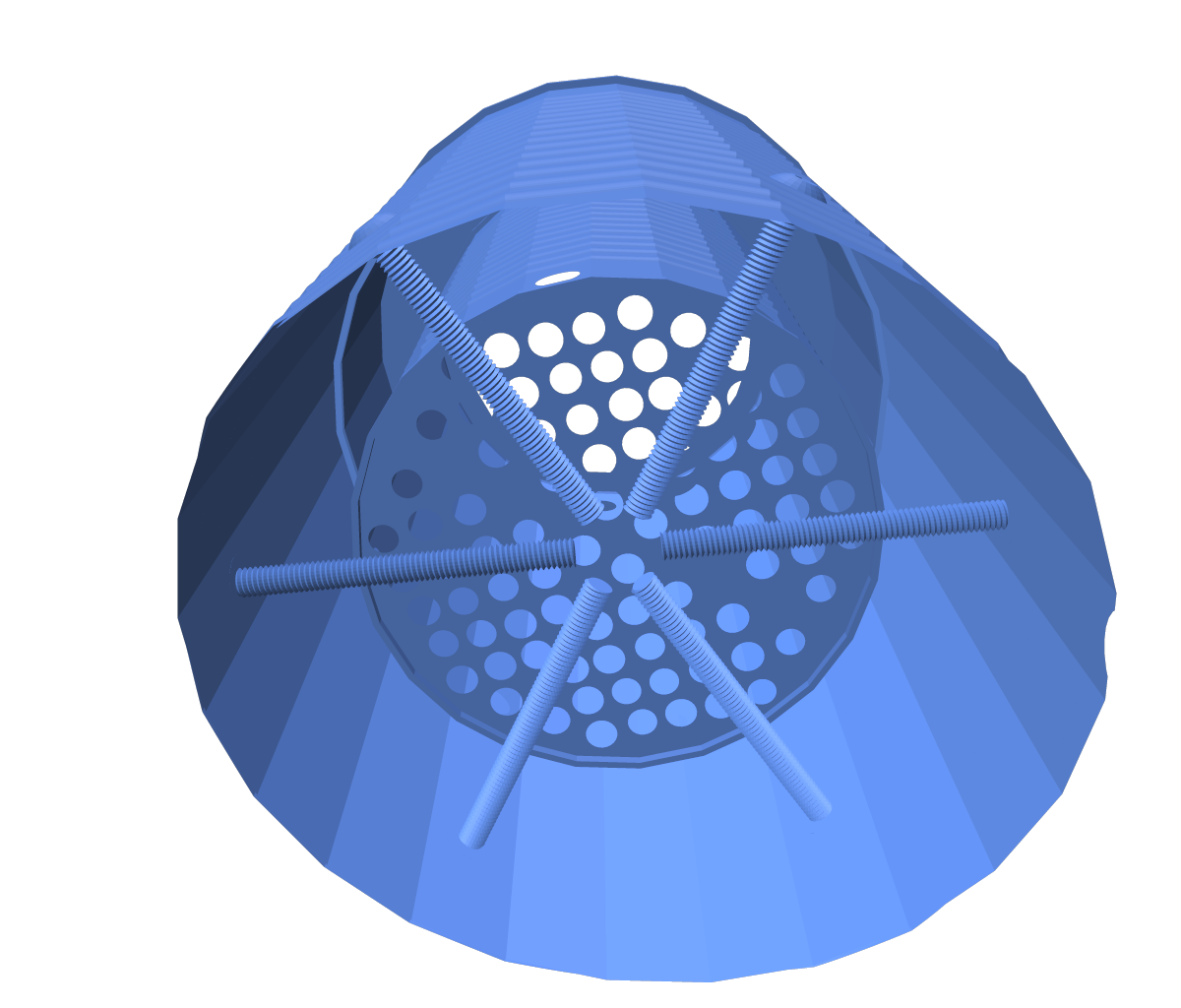}};
    \spy on (0.2,-0.1) in node [left] at (-0.4,0.25);
\end{tikzpicture} & \begin{tikzpicture}[spy using outlines={circle,yellow,magnification=4,size=2.0cm, connect spies}]
    \node {\includegraphics[width=0.1\textwidth]{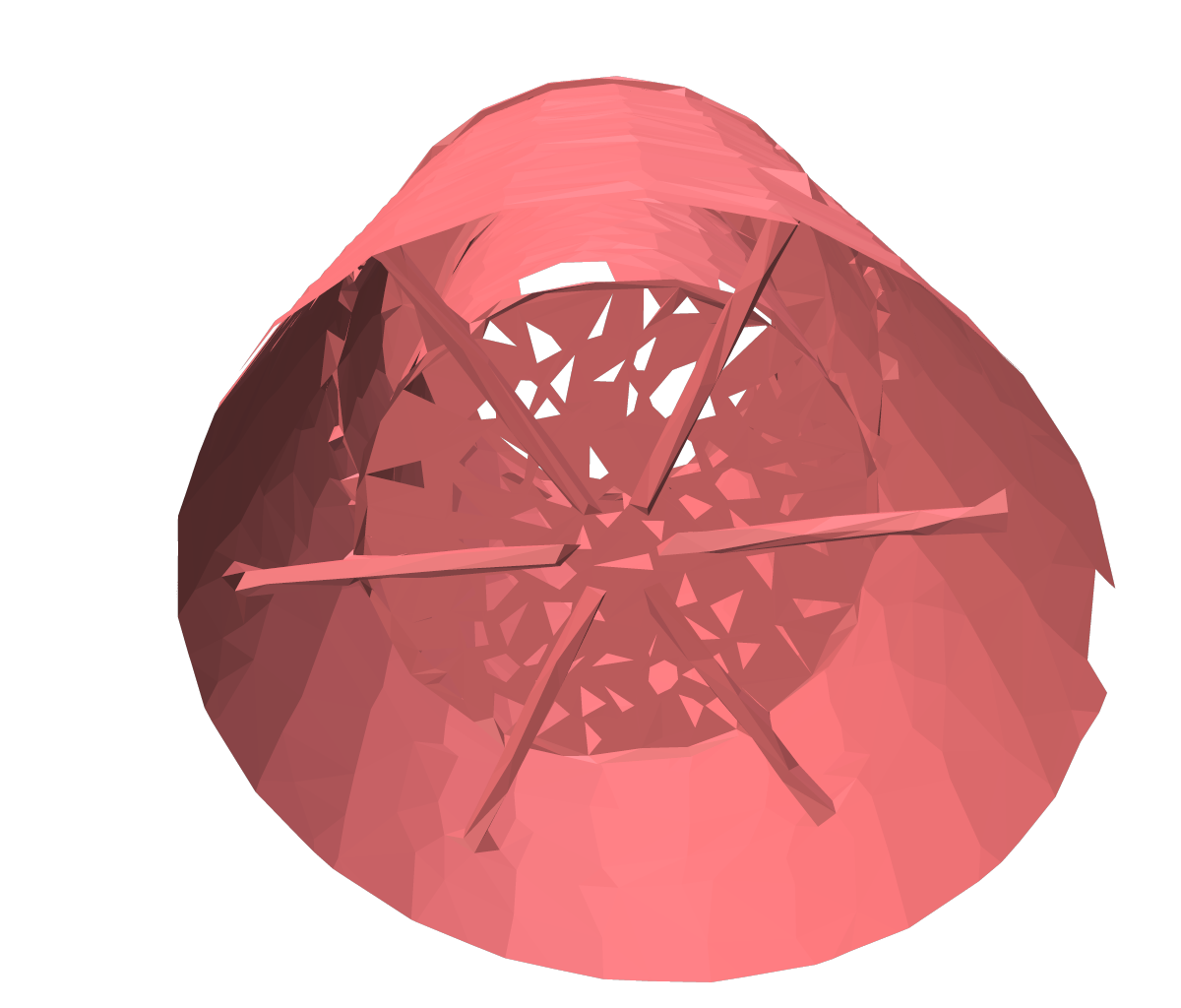}};
    \spy on (0.2,-0.1) in node [left] at (-0.4,0.25);
\end{tikzpicture} \\
\begin{tikzpicture}[spy using outlines={circle,yellow,magnification=3,size=2.0cm, connect spies}]
    \node {\includegraphics[width=0.1\textwidth]{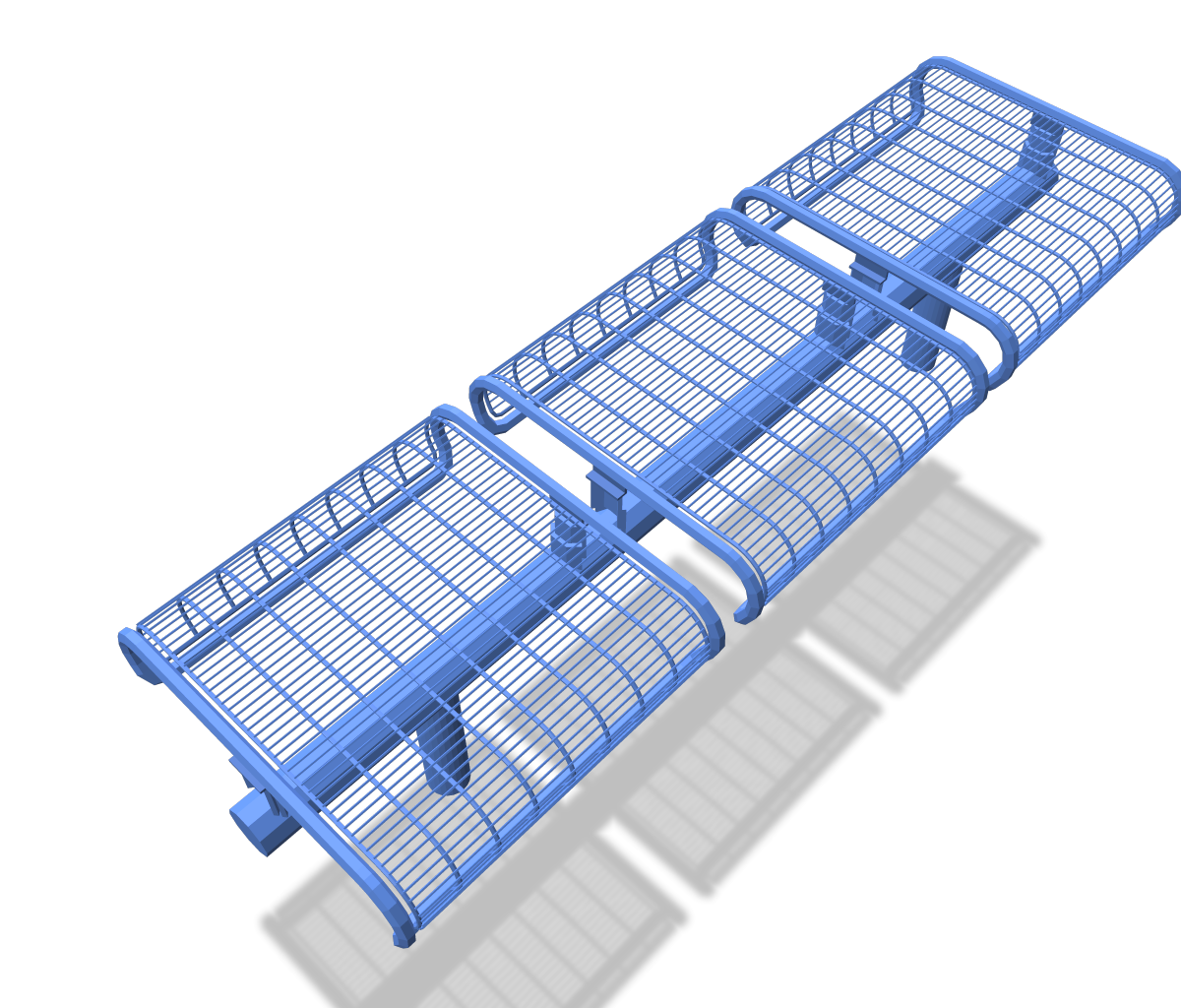}};
    \spy on (-0.4,-0.3) in node [left] at (-0.4,0.25);
\end{tikzpicture} & \begin{tikzpicture}[spy using outlines={circle,yellow,magnification=3,size=2.0cm, connect spies}]
    \node {\includegraphics[width=0.1\textwidth]{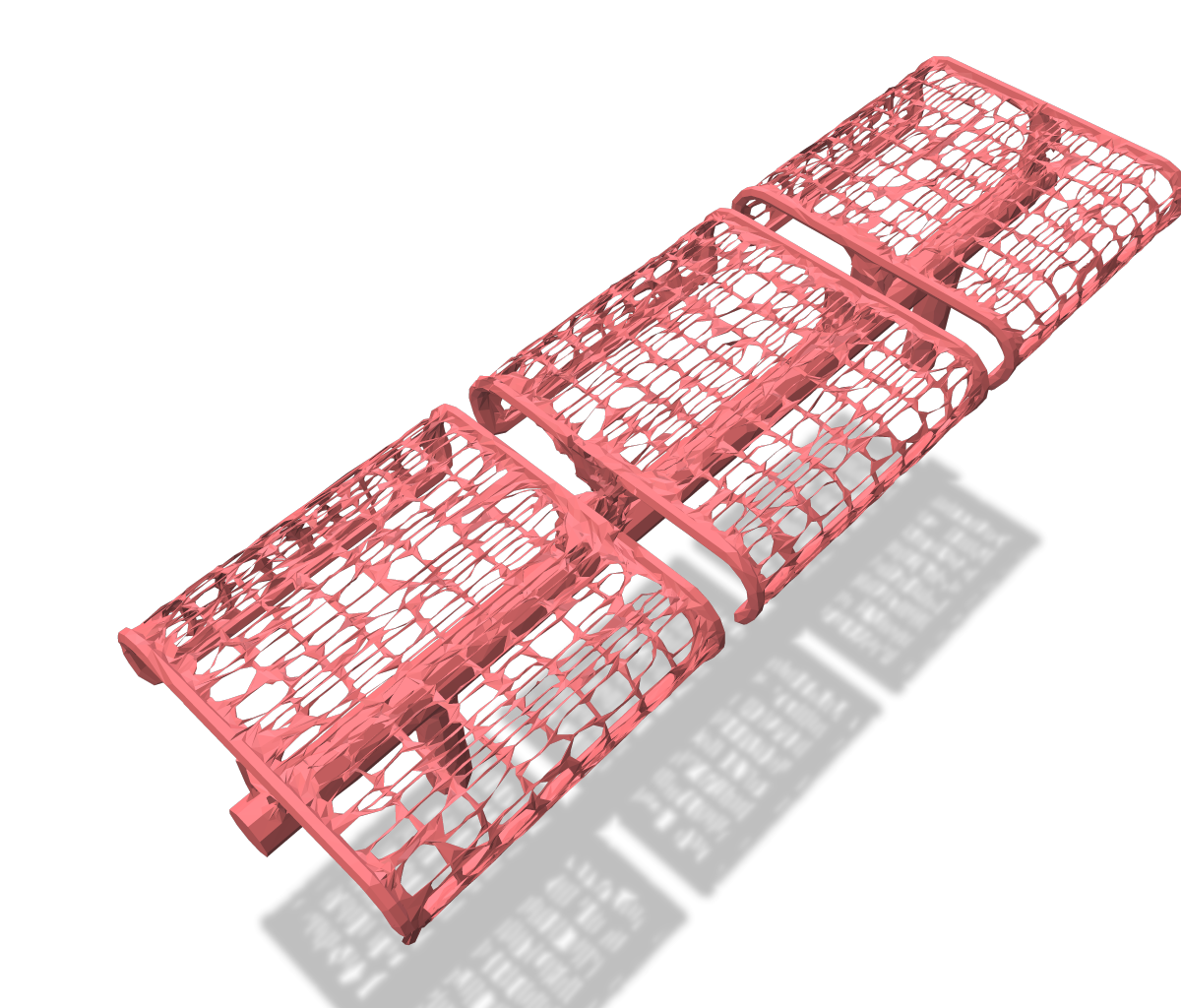}};
    \spy on (-0.4,-0.3) in node [left] at (-0.4,0.25);
\end{tikzpicture}

\end{tabular}

\caption{Limitations. The target is on the left. Top: the inner structure of a can was poorly optimized (holes and rods) because it cannot be seen from the outside, which means no render loss gradients are propagated to the vertex locations. Bottom: the fine structure of the bench mesh is not preserved in the output, due to initialization failing to capture its genus correctly.}
\label{fig:fails}
\end{figure}

\subsection{Limitations and Future Work}

Typical for all render based techniques, inner structures are overall less accurate than visible ones (\Cref{fig:fails}, Top). We partially negate this problem by using a 3D loss term, but it works best when combined with a render loss, and thus the final inner structures are not as accurate. This could possibly be improved by cleverly rendering from within enclosed spaces or by depth peeling.

Additionally, leveraging a volume based initialization has one significant drawback, where very thin structured are skipped. Since our method does not alter the topology of the initial source mesh, this leads to artifacts in the reconstruction (\Cref{fig:fails}, Bottom). This may be improved by leveraging a different initialization, but warrants further investigation for schemes that lend themselves well to render-based optimization.

Lastly, since we used PyTorch \cite{pytorch} to implement every step of our pipeline, topological operations are differentiable with respect to the vertices locations. This means back-propagating through them is possible and was demonstrated by our use of our projection both as a geometric predicate and a loss. Ultimately, this property can be used to decide where faces should be split using a learning method. We hope to leverage this for learning over topological operations jointly with geometry optimization, as we believe this is key to improve deep learning applications over meshes.
\section{Acknowledgments}
This work was partially supported by Len Blavatnik and the Blavatnik family foundation, the Yandex Initiative in Machine Learning, ISF (number 1337/22), and BSF (number 2020280)
\bibliographystyle{ACM-Reference-Format}
\bibliography{bibliography}


\begin{thebibliography}{39}


\ifx \showCODEN    \undefined \def \showCODEN     #1{\unskip}     \fi
\ifx \showDOI      \undefined \def \showDOI       #1{#1}\fi
\ifx \showISBNx    \undefined \def \showISBNx     #1{\unskip}     \fi
\ifx \showISBNxiii \undefined \def \showISBNxiii  #1{\unskip}     \fi
\ifx \showISSN     \undefined \def \showISSN      #1{\unskip}     \fi
\ifx \showLCCN     \undefined \def \showLCCN      #1{\unskip}     \fi
\ifx \shownote     \undefined \def \shownote      #1{#1}          \fi
\ifx \showarticletitle \undefined \def \showarticletitle #1{#1}   \fi
\ifx \showURL      \undefined \def \showURL       {\relax}        \fi
\providecommand\bibfield[2]{#2}
\providecommand\bibinfo[2]{#2}
\providecommand\natexlab[1]{#1}
\providecommand\showeprint[2][]{arXiv:#2}

\bibitem[Alliez et~al\mbox{.}(2023)]%
        {alphawrap_cgal}
\bibfield{author}{\bibinfo{person}{Pierre Alliez}, \bibinfo{person}{David
  Cohen-Steiner}, \bibinfo{person}{Michael Hemmer}, \bibinfo{person}{C{\'e}dric
  Portaneri}, {and} \bibinfo{person}{Mael Rouxel-Labb{\'e}}.}
  \bibinfo{year}{2023}\natexlab{}.
\newblock \showarticletitle{{3D} Alpha Wrapping}.
\newblock In \bibinfo{booktitle}{\emph{{CGAL} User and Reference Manual}
  (\bibinfo{edition}{{5.5.2}} ed.)}. \bibinfo{publisher}{{CGAL Editorial
  Board}}.
\newblock
\urldef\tempurl%
\url{https://doc.cgal.org/5.5.2/Manual/packages.html#PkgAlphaWrap3}
\showURL{%
\tempurl}


\bibitem[Aspert et~al\mbox{.}(2002)]%
        {supersample_mesh}
\bibfield{author}{\bibinfo{person}{Nicolas Aspert}, \bibinfo{person}{Diego
  Santa-Cruz}, {and} \bibinfo{person}{Touradj Ebrahimi}.}
  \bibinfo{year}{2002}\natexlab{}.
\newblock \showarticletitle{Mesh: Measuring errors between surfaces using the
  hausdorff distance}. In \bibinfo{booktitle}{\emph{Proceedings. IEEE
  international conference on multimedia and expo}}, Vol.~\bibinfo{volume}{1}.
  IEEE, \bibinfo{pages}{705--708}.
\newblock


\bibitem[Chang et~al\mbox{.}(2015)]%
        {shapenet2015}
\bibfield{author}{\bibinfo{person}{Angel~X. Chang}, \bibinfo{person}{Thomas
  Funkhouser}, \bibinfo{person}{Leonidas Guibas}, \bibinfo{person}{Pat
  Hanrahan}, \bibinfo{person}{Qixing Huang}, \bibinfo{person}{Zimo Li},
  \bibinfo{person}{Silvio Savarese}, \bibinfo{person}{Manolis Savva},
  \bibinfo{person}{Shuran Song}, \bibinfo{person}{Hao Su},
  \bibinfo{person}{Jianxiong Xiao}, \bibinfo{person}{Li Yi}, {and}
  \bibinfo{person}{Fisher Yu}.} \bibinfo{year}{2015}\natexlab{}.
\newblock \bibinfo{booktitle}{\emph{{ShapeNet: An Information-Rich 3D Model
  Repository}}}.
\newblock \bibinfo{type}{{T}echnical {R}eport} arXiv:1512.03012 [cs.GR].
  \bibinfo{institution}{Stanford University --- Princeton University --- Toyota
  Technological Institute at Chicago}.
\newblock


\bibitem[Chu et~al\mbox{.}(2019)]%
        {MeshRepair}
\bibfield{author}{\bibinfo{person}{Lei Chu}, \bibinfo{person}{Hao Pan},
  \bibinfo{person}{Yang Liu}, {and} \bibinfo{person}{Wenping Wang}.}
  \bibinfo{year}{2019}\natexlab{}.
\newblock \showarticletitle{Repairing Man-Made Meshes via Visual Driven Global
  Optimization with Minimum Intrusion}.
\newblock \bibinfo{journal}{\emph{ACM Trans. Graph. (SIGGRAPH ASIA)}}
  \bibinfo{volume}{38}, \bibinfo{number}{6} (\bibinfo{year}{2019}),
  \bibinfo{pages}{158:1--158:18}.
\newblock
\urldef\tempurl%
\url{https://doi.org/10.1145/3355089.3356507}
\showDOI{\tempurl}


\bibitem[Cignoni et~al\mbox{.}(2008)]%
        {meshlab}
\bibfield{author}{\bibinfo{person}{Paolo Cignoni}, \bibinfo{person}{Marco
  Callieri}, \bibinfo{person}{Massimiliano Corsini}, \bibinfo{person}{Matteo
  Dellepiane}, \bibinfo{person}{Fabio Ganovelli}, {and} \bibinfo{person}{Guido
  Ranzuglia}.} \bibinfo{year}{2008}\natexlab{}.
\newblock \showarticletitle{{MeshLab: an Open-Source Mesh Processing Tool}}. In
  \bibinfo{booktitle}{\emph{Eurographics Italian Chapter Conference}},
  \bibfield{editor}{\bibinfo{person}{Vittorio Scarano},
  \bibinfo{person}{Rosario~De Chiara}, {and} \bibinfo{person}{Ugo Erra}}
  (Eds.). \bibinfo{publisher}{The Eurographics Association}.
\newblock
\showISBNx{978-3-905673-68-5}
\urldef\tempurl%
\url{https://doi.org/10.2312/LocalChapterEvents/ItalChap/ItalianChapConf2008/129-136}
\showDOI{\tempurl}


\bibitem[Cohen-Steiner and Da(2004)]%
        {advancing_fronts_cgal}
\bibfield{author}{\bibinfo{person}{David Cohen-Steiner} {and}
  \bibinfo{person}{Frank Da}.} \bibinfo{year}{2004}\natexlab{}.
\newblock \showarticletitle{A greedy Delaunay-based surface reconstruction
  algorithm}.
\newblock \bibinfo{journal}{\emph{The visual computer}}  \bibinfo{volume}{20}
  (\bibinfo{year}{2004}), \bibinfo{pages}{4--16}.
\newblock


\bibitem[Deitke et~al\mbox{.}(2022)]%
        {deitke2022objaverse}
\bibfield{author}{\bibinfo{person}{Matt Deitke}, \bibinfo{person}{Dustin
  Schwenk}, \bibinfo{person}{Jordi Salvador}, \bibinfo{person}{Luca Weihs},
  \bibinfo{person}{Oscar Michel}, \bibinfo{person}{Eli VanderBilt},
  \bibinfo{person}{Ludwig Schmidt}, \bibinfo{person}{Kiana Ehsani},
  \bibinfo{person}{Aniruddha Kembhavi}, {and} \bibinfo{person}{Ali Farhadi}.}
  \bibinfo{year}{2022}\natexlab{}.
\newblock \showarticletitle{Objaverse: A Universe of Annotated 3D Objects}.
\newblock \bibinfo{journal}{\emph{arXiv preprint arXiv:2212.08051}}
  (\bibinfo{year}{2022}).
\newblock


\bibitem[Dunyach et~al\mbox{.}(2013)]%
        {ada_remesh}
\bibfield{author}{\bibinfo{person}{Marion Dunyach}, \bibinfo{person}{David
  Vanderhaeghe}, \bibinfo{person}{Loïc Barthe}, {and} \bibinfo{person}{Mario
  Botsch}.} \bibinfo{year}{2013}\natexlab{}.
\newblock \showarticletitle{{Adaptive Remeshing for Real-Time Mesh
  Deformation}}. In \bibinfo{booktitle}{\emph{Eurographics 2013 - Short
  Papers}}, \bibfield{editor}{\bibinfo{person}{M.-A. Otaduy} {and}
  \bibinfo{person}{O.~Sorkine}} (Eds.). \bibinfo{publisher}{The Eurographics
  Association}.
\newblock
\showISSN{1017-4656}
\urldef\tempurl%
\url{https://doi.org/10.2312/conf/EG2013/short/029-032}
\showDOI{\tempurl}


\bibitem[Gao et~al\mbox{.}(2022)]%
        {gao2022get3d}
\bibfield{author}{\bibinfo{person}{Jun Gao}, \bibinfo{person}{Tianchang Shen},
  \bibinfo{person}{Zian Wang}, \bibinfo{person}{Wenzheng Chen},
  \bibinfo{person}{Kangxue Yin}, \bibinfo{person}{Daiqing Li},
  \bibinfo{person}{Or Litany}, \bibinfo{person}{Zan Gojcic}, {and}
  \bibinfo{person}{Sanja Fidler}.} \bibinfo{year}{2022}\natexlab{}.
\newblock \showarticletitle{GET3D: A Generative Model of High Quality 3D
  Textured Shapes Learned from Images}. In \bibinfo{booktitle}{\emph{Advances
  In Neural Information Processing Systems}}.
\newblock


\bibitem[Garland and Heckbert(1997)]%
        {qslim}
\bibfield{author}{\bibinfo{person}{Michael Garland} {and} \bibinfo{person}{Paul
  Heckbert}.} \bibinfo{year}{1997}\natexlab{}.
\newblock \showarticletitle{Surface Simplification Using Quadric Error
  Metrics}.
\newblock \bibinfo{journal}{\emph{Proceedings of the ACM SIGGRAPH Conference on
  Computer Graphics}}  \bibinfo{volume}{1997} (\bibinfo{date}{07}
  \bibinfo{year}{1997}).
\newblock
\urldef\tempurl%
\url{https://doi.org/10.1145/258734.258849}
\showDOI{\tempurl}


\bibitem[Hanocka et~al\mbox{.}(2019)]%
        {meshcnn}
\bibfield{author}{\bibinfo{person}{Rana Hanocka}, \bibinfo{person}{Amir Hertz},
  \bibinfo{person}{Noa Fish}, \bibinfo{person}{Raja Giryes},
  \bibinfo{person}{Shachar Fleishman}, {and} \bibinfo{person}{Daniel
  Cohen-Or}.} \bibinfo{year}{2019}\natexlab{}.
\newblock \showarticletitle{MeshCNN: A Network with an Edge}.
\newblock \bibinfo{journal}{\emph{ACM Transactions on Graphics (TOG)}}
  \bibinfo{volume}{38}, \bibinfo{number}{4} (\bibinfo{year}{2019}),
  \bibinfo{pages}{90}.
\newblock


\bibitem[Hanocka et~al\mbox{.}(2020)]%
        {point2mesh}
\bibfield{author}{\bibinfo{person}{Rana Hanocka}, \bibinfo{person}{Gal Metzer},
  \bibinfo{person}{Raja Giryes}, {and} \bibinfo{person}{Daniel Cohen-Or}.}
  \bibinfo{year}{2020}\natexlab{}.
\newblock \showarticletitle{Point2Mesh: A Self-Prior for Deformable Meshes}.
\newblock \bibinfo{journal}{\emph{ACM Trans. Graph.}} \bibinfo{volume}{39},
  \bibinfo{number}{4}, Article \bibinfo{articleno}{126} (\bibinfo{date}{aug}
  \bibinfo{year}{2020}), \bibinfo{numpages}{12}~pages.
\newblock
\showISSN{0730-0301}
\urldef\tempurl%
\url{https://doi.org/10.1145/3386569.3392415}
\showDOI{\tempurl}


\bibitem[Hoppe et~al\mbox{.}(1993)]%
        {hoppe}
\bibfield{author}{\bibinfo{person}{Hugues Hoppe}, \bibinfo{person}{Tony
  DeRose}, \bibinfo{person}{Tom Duchamp}, \bibinfo{person}{John McDonald},
  {and} \bibinfo{person}{Werner Stuetzle}.} \bibinfo{year}{1993}\natexlab{}.
\newblock \showarticletitle{Mesh optimization}. In
  \bibinfo{booktitle}{\emph{Proceedings of the 20th annual conference on
  Computer graphics and interactive techniques}}. \bibinfo{pages}{19--26}.
\newblock


\bibitem[Hu and Chen(2019)]%
        {chamfer_self_intersect}
\bibfield{author}{\bibinfo{person}{Siyu Hu} {and} \bibinfo{person}{Xuejin
  Chen}.} \bibinfo{year}{2019}\natexlab{}.
\newblock \showarticletitle{Preventing self-intersection with cycle
  regularization in neural networks for mesh reconstruction from a single RGB
  image}.
\newblock \bibinfo{journal}{\emph{Computer Aided Geometric Design}}
  \bibinfo{volume}{72} (\bibinfo{year}{2019}), \bibinfo{pages}{84--97}.
\newblock


\bibitem[Hu et~al\mbox{.}(2020)]%
        {ftetwild}
\bibfield{author}{\bibinfo{person}{Yixin Hu}, \bibinfo{person}{Teseo
  Schneider}, \bibinfo{person}{Bolun Wang}, \bibinfo{person}{Denis Zorin},
  {and} \bibinfo{person}{Daniele Panozzo}.} \bibinfo{year}{2020}\natexlab{}.
\newblock \showarticletitle{Fast Tetrahedral Meshing in the Wild}.
\newblock \bibinfo{journal}{\emph{ACM Trans. Graph.}} \bibinfo{volume}{39},
  \bibinfo{number}{4}, Article \bibinfo{articleno}{117} (\bibinfo{date}{July}
  \bibinfo{year}{2020}), \bibinfo{numpages}{18}~pages.
\newblock
\showISSN{0730-0301}
\urldef\tempurl%
\url{https://doi.org/10.1145/3386569.3392385}
\showDOI{\tempurl}


\bibitem[Hu et~al\mbox{.}(2018)]%
        {tetwild}
\bibfield{author}{\bibinfo{person}{Yixin Hu}, \bibinfo{person}{Qingnan Zhou},
  \bibinfo{person}{Xifeng Gao}, \bibinfo{person}{Alec Jacobson},
  \bibinfo{person}{Denis Zorin}, {and} \bibinfo{person}{Daniele Panozzo}.}
  \bibinfo{year}{2018}\natexlab{}.
\newblock \showarticletitle{Tetrahedral Meshing in the Wild}.
\newblock \bibinfo{journal}{\emph{ACM Trans. Graph.}} \bibinfo{volume}{37},
  \bibinfo{number}{4}, Article \bibinfo{articleno}{60} (\bibinfo{date}{July}
  \bibinfo{year}{2018}), \bibinfo{numpages}{14}~pages.
\newblock
\showISSN{0730-0301}
\urldef\tempurl%
\url{https://doi.org/10.1145/3197517.3201353}
\showDOI{\tempurl}


\bibitem[Huang et~al\mbox{.}(2018)]%
        {manifold}
\bibfield{author}{\bibinfo{person}{Jingwei Huang}, \bibinfo{person}{Hao Su},
  {and} \bibinfo{person}{Leonidas Guibas}.} \bibinfo{year}{2018}\natexlab{}.
\newblock \showarticletitle{Robust watertight manifold surface generation
  method for shapenet models}.
\newblock \bibinfo{journal}{\emph{arXiv preprint arXiv:1802.01698}}
  (\bibinfo{year}{2018}).
\newblock


\bibitem[Huang et~al\mbox{.}(2020)]%
        {manifold_plus}
\bibfield{author}{\bibinfo{person}{Jingwei Huang}, \bibinfo{person}{Yichao
  Zhou}, {and} \bibinfo{person}{Leonidas Guibas}.}
  \bibinfo{year}{2020}\natexlab{}.
\newblock \showarticletitle{ManifoldPlus: A Robust and Scalable Watertight
  Manifold Surface Generation Method for Triangle Soups}.
\newblock \bibinfo{journal}{\emph{arXiv preprint arXiv:2005.11621}}
  (\bibinfo{year}{2020}).
\newblock


\bibitem[Kazhdan et~al\mbox{.}(2006)]%
        {poisson}
\bibfield{author}{\bibinfo{person}{Michael Kazhdan}, \bibinfo{person}{Matthew
  Bolitho}, {and} \bibinfo{person}{Hugues Hoppe}.}
  \bibinfo{year}{2006}\natexlab{}.
\newblock \showarticletitle{Poisson surface reconstruction}. In
  \bibinfo{booktitle}{\emph{Proceedings of the fourth Eurographics symposium on
  Geometry processing}}, Vol.~\bibinfo{volume}{7}.
\newblock


\bibitem[Kazhdan and Hoppe(2013)]%
        {screened_poisson}
\bibfield{author}{\bibinfo{person}{Michael Kazhdan} {and}
  \bibinfo{person}{Hugues Hoppe}.} \bibinfo{year}{2013}\natexlab{}.
\newblock \showarticletitle{Screened poisson surface reconstruction}.
\newblock \bibinfo{journal}{\emph{ACM Transactions on Graphics (TOG)}}
  \bibinfo{volume}{32}, \bibinfo{number}{3} (\bibinfo{year}{2013}),
  \bibinfo{pages}{29}.
\newblock


\bibitem[Kobbelt(2000)]%
        {sqrt3}
\bibfield{author}{\bibinfo{person}{Leif Kobbelt}.}
  \bibinfo{year}{2000}\natexlab{}.
\newblock \showarticletitle{sqrt3-Subdivision}. In
  \bibinfo{booktitle}{\emph{Proceedings of the 27th Annual Conference on
  Computer Graphics and Interactive Techniques}}
  \emph{(\bibinfo{series}{SIGGRAPH '00})}. \bibinfo{publisher}{ACM
  Press/Addison-Wesley Publishing Co.}, \bibinfo{address}{USA},
  \bibinfo{pages}{103–112}.
\newblock
\showISBNx{1581132085}
\urldef\tempurl%
\url{https://doi.org/10.1145/344779.344835}
\showDOI{\tempurl}


\bibitem[Krishnamurthy and Levoy(1996)]%
        {armadillo}
\bibfield{author}{\bibinfo{person}{Venkat Krishnamurthy} {and}
  \bibinfo{person}{Marc Levoy}.} \bibinfo{year}{1996}\natexlab{}.
\newblock \showarticletitle{Fitting smooth surfaces to dense polygon meshes}.
  In \bibinfo{booktitle}{\emph{Proceedings of the 23rd annual conference on
  Computer graphics and interactive techniques}}. \bibinfo{pages}{313--324}.
\newblock


\bibitem[Laine et~al\mbox{.}(2020)]%
        {nvdiffrast}
\bibfield{author}{\bibinfo{person}{Samuli Laine}, \bibinfo{person}{Janne
  Hellsten}, \bibinfo{person}{Tero Karras}, \bibinfo{person}{Yeongho Seol},
  \bibinfo{person}{Jaakko Lehtinen}, {and} \bibinfo{person}{Timo Aila}.}
  \bibinfo{year}{2020}\natexlab{}.
\newblock \showarticletitle{Modular Primitives for High-Performance
  Differentiable Rendering}.
\newblock \bibinfo{journal}{\emph{ACM Trans. Graph.}} \bibinfo{volume}{39},
  \bibinfo{number}{6}, Article \bibinfo{articleno}{194} (\bibinfo{date}{nov}
  \bibinfo{year}{2020}), \bibinfo{numpages}{14}~pages.
\newblock
\showISSN{0730-0301}
\urldef\tempurl%
\url{https://doi.org/10.1145/3414685.3417861}
\showDOI{\tempurl}


\bibitem[Ling et~al\mbox{.}(2022)]%
        {ling2022vectoradam}
\bibfield{author}{\bibinfo{person}{Selena Ling}, \bibinfo{person}{Nicholas
  Sharp}, {and} \bibinfo{person}{Alec Jacobson}.}
  \bibinfo{year}{2022}\natexlab{}.
\newblock \bibinfo{title}{VectorAdam for Rotation Equivariant Geometry
  Optimization}.
\newblock
\newblock
\showeprint[arxiv]{2205.13599}~[cs.LG]


\bibitem[Low(2004)]%
        {planar_icp}
\bibfield{author}{\bibinfo{person}{Kok-Lim Low}.}
  \bibinfo{year}{2004}\natexlab{}.
\newblock \showarticletitle{Linear least-squares optimization for
  point-to-plane icp surface registration}.
\newblock \bibinfo{journal}{\emph{Chapel Hill, University of North Carolina}}
  \bibinfo{volume}{4}, \bibinfo{number}{10} (\bibinfo{year}{2004}),
  \bibinfo{pages}{1--3}.
\newblock


\bibitem[Milano et~al\mbox{.}(2020)]%
        {pd_mesh_cnn}
\bibfield{author}{\bibinfo{person}{Francesco Milano}, \bibinfo{person}{Antonio
  Loquercio}, \bibinfo{person}{Antoni Rosinol}, \bibinfo{person}{Davide
  Scaramuzza}, {and} \bibinfo{person}{Luca Carlone}.}
  \bibinfo{year}{2020}\natexlab{}.
\newblock \showarticletitle{Primal-Dual Mesh Convolutional Neural Networks}. In
  \bibinfo{booktitle}{\emph{Advances in Neural Information Processing
  Systems}}, \bibfield{editor}{\bibinfo{person}{H.~Larochelle},
  \bibinfo{person}{M.~Ranzato}, \bibinfo{person}{R.~Hadsell},
  \bibinfo{person}{M.~F. Balcan}, {and} \bibinfo{person}{H.~Lin}} (Eds.),
  Vol.~\bibinfo{volume}{33}. \bibinfo{publisher}{Curran Associates, Inc.},
  \bibinfo{pages}{952--963}.
\newblock
\urldef\tempurl%
\url{https://proceedings.neurips.cc/paper/2020/file/0a656cc19f3f5b41530182a9e03982a4-Paper.pdf}
\showURL{%
\tempurl}


\bibitem[Nicolet et~al\mbox{.}(2021)]%
        {large_steps}
\bibfield{author}{\bibinfo{person}{Baptiste Nicolet}, \bibinfo{person}{Alec
  Jacobson}, {and} \bibinfo{person}{Wenzel Jakob}.}
  \bibinfo{year}{2021}\natexlab{}.
\newblock \showarticletitle{Large Steps in Inverse Rendering of Geometry}.
\newblock \bibinfo{journal}{\emph{ACM Trans. Graph.}} \bibinfo{volume}{40},
  \bibinfo{number}{6}, Article \bibinfo{articleno}{248} (\bibinfo{date}{dec}
  \bibinfo{year}{2021}), \bibinfo{numpages}{13}~pages.
\newblock
\showISSN{0730-0301}
\urldef\tempurl%
\url{https://doi.org/10.1145/3478513.3480501}
\showDOI{\tempurl}


\bibitem[{\"O}ztireli et~al\mbox{.}(2009)]%
        {RIMLS_marching}
\bibfield{author}{\bibinfo{person}{A~Cengiz {\"O}ztireli},
  \bibinfo{person}{Gael Guennebaud}, {and} \bibinfo{person}{Markus Gross}.}
  \bibinfo{year}{2009}\natexlab{}.
\newblock \showarticletitle{Feature preserving point set surfaces based on
  non-linear kernel regression}. In \bibinfo{booktitle}{\emph{Computer graphics
  forum}}, Vol.~\bibinfo{volume}{28}. Wiley Online Library,
  \bibinfo{pages}{493--501}.
\newblock


\bibitem[Palfinger(2022)]%
        {palfinger2022continuous}
\bibfield{author}{\bibinfo{person}{Werner Palfinger}.}
  \bibinfo{year}{2022}\natexlab{}.
\newblock \showarticletitle{Continuous remeshing for inverse rendering}.
\newblock \bibinfo{journal}{\emph{Computer Animation and Virtual Worlds}}
  (\bibinfo{year}{2022}), \bibinfo{pages}{e2101}.
\newblock


\bibitem[Paszke et~al\mbox{.}(2019)]%
        {pytorch}
\bibfield{author}{\bibinfo{person}{Adam Paszke}, \bibinfo{person}{Sam Gross},
  \bibinfo{person}{Francisco Massa}, \bibinfo{person}{Adam Lerer},
  \bibinfo{person}{James Bradbury}, \bibinfo{person}{Gregory Chanan},
  \bibinfo{person}{Trevor Killeen}, \bibinfo{person}{Zeming Lin},
  \bibinfo{person}{Natalia Gimelshein}, \bibinfo{person}{Luca Antiga},
  \bibinfo{person}{Alban Desmaison}, \bibinfo{person}{Andreas Kopf},
  \bibinfo{person}{Edward Yang}, \bibinfo{person}{Zachary DeVito},
  \bibinfo{person}{Martin Raison}, \bibinfo{person}{Alykhan Tejani},
  \bibinfo{person}{Sasank Chilamkurthy}, \bibinfo{person}{Benoit Steiner},
  \bibinfo{person}{Lu Fang}, \bibinfo{person}{Junjie Bai}, {and}
  \bibinfo{person}{Soumith Chintala}.} \bibinfo{year}{2019}\natexlab{}.
\newblock \showarticletitle{PyTorch: An Imperative Style, High-Performance Deep
  Learning Library}.
\newblock In \bibinfo{booktitle}{\emph{Advances in Neural Information
  Processing Systems 32}}, \bibfield{editor}{\bibinfo{person}{H.~Wallach},
  \bibinfo{person}{H.~Larochelle}, \bibinfo{person}{A.~Beygelzimer},
  \bibinfo{person}{F.~d\textquotesingle Alch\'{e}-Buc},
  \bibinfo{person}{E.~Fox}, {and} \bibinfo{person}{R.~Garnett}} (Eds.).
  \bibinfo{publisher}{Curran Associates, Inc.}, \bibinfo{pages}{8024--8035}.
\newblock
\urldef\tempurl%
\url{http://papers.neurips.cc/paper/9015-pytorch-an-imperative-style-high-performance-deep-learning-library.pdf}
\showURL{%
\tempurl}


\bibitem[Peng et~al\mbox{.}(2021)]%
        {shape_as_points}
\bibfield{author}{\bibinfo{person}{Songyou Peng}, \bibinfo{person}{Chiyu~"Max"
  Jiang}, \bibinfo{person}{Yiyi Liao}, \bibinfo{person}{Michael Niemeyer},
  \bibinfo{person}{Marc Pollefeys}, {and} \bibinfo{person}{Andreas Geiger}.}
  \bibinfo{year}{2021}\natexlab{}.
\newblock \showarticletitle{Shape As Points: A Differentiable Poisson Solver}.
  In \bibinfo{booktitle}{\emph{Advances in Neural Information Processing
  Systems (NeurIPS)}}.
\newblock


\bibitem[Sharf et~al\mbox{.}(2006)]%
        {competing_fronts}
\bibfield{author}{\bibinfo{person}{Andrei Sharf}, \bibinfo{person}{Thomas
  Lewiner}, \bibinfo{person}{Ariel Shamir}, \bibinfo{person}{Leif Kobbelt},
  {and} \bibinfo{person}{Daniel Cohen-Or}.} \bibinfo{year}{2006}\natexlab{}.
\newblock \showarticletitle{Competing fronts for coarse--to--fine surface
  reconstruction}. In \bibinfo{booktitle}{\emph{Computer Graphics Forum}},
  Vol.~\bibinfo{volume}{25}. Wiley Online Library, \bibinfo{pages}{389--398}.
\newblock


\bibitem[Smirnov and Solomon(2021)]%
        {hodgenet}
\bibfield{author}{\bibinfo{person}{Dmitriy Smirnov} {and}
  \bibinfo{person}{Justin Solomon}.} \bibinfo{year}{2021}\natexlab{}.
\newblock \bibinfo{title}{HodgeNet: Learning Spectral Geometry on Triangle
  Meshes}.
\newblock
\newblock
\showeprint[arxiv]{2104.12826}~[cs.GR]


\bibitem[Takayama et~al\mbox{.}(2014)]%
        {cleanmesh}
\bibfield{author}{\bibinfo{person}{Kenshi Takayama}, \bibinfo{person}{Alec
  Jacobson}, \bibinfo{person}{Ladislav Kavan}, {and} \bibinfo{person}{Olga
  Sorkine-Hornung}.} \bibinfo{year}{2014}\natexlab{}.
\newblock \showarticletitle{A Simple Method for Correcting Facet Orientations
  in Polygon Meshes Based on Ray Casting}.
\newblock \bibinfo{journal}{\emph{Journal of Computer Graphics Techniques
  (JCGT)}} \bibinfo{volume}{3}, \bibinfo{number}{4} (\bibinfo{date}{19
  December} \bibinfo{year}{2014}), \bibinfo{pages}{53--63}.
\newblock
\showISSN{2331-7418}
\urldef\tempurl%
\url{http://jcgt.org/published/0003/04/02/}
\showURL{%
\tempurl}


\bibitem[Turk and Levoy(1994)]%
        {bunny}
\bibfield{author}{\bibinfo{person}{Greg Turk} {and} \bibinfo{person}{Marc
  Levoy}.} \bibinfo{year}{1994}\natexlab{}.
\newblock \showarticletitle{Zippered polygon meshes from range images}. In
  \bibinfo{booktitle}{\emph{Proceedings of the 21st annual conference on
  Computer graphics and interactive techniques}}. \bibinfo{pages}{311--318}.
\newblock


\bibitem[Wu et~al\mbox{.}(2015)]%
        {modelnet}
\bibfield{author}{\bibinfo{person}{Zhirong Wu}, \bibinfo{person}{Shuran Song},
  \bibinfo{person}{Aditya Khosla}, \bibinfo{person}{Fisher Yu},
  \bibinfo{person}{Linguang Zhang}, \bibinfo{person}{Xiaoou Tang}, {and}
  \bibinfo{person}{Jianxiong Xiao}.} \bibinfo{year}{2015}\natexlab{}.
\newblock \showarticletitle{3d shapenets: A deep representation for volumetric
  shapes}. In \bibinfo{booktitle}{\emph{Proceedings of the IEEE conference on
  computer vision and pattern recognition}}. \bibinfo{pages}{1912--1920}.
\newblock


\bibitem[Yao et~al\mbox{.}(2020)]%
        {blendedMVS}
\bibfield{author}{\bibinfo{person}{Yao Yao}, \bibinfo{person}{Zixin Luo},
  \bibinfo{person}{Shiwei Li}, \bibinfo{person}{Jingyang Zhang},
  \bibinfo{person}{Yufan Ren}, \bibinfo{person}{Lei Zhou},
  \bibinfo{person}{Tian Fang}, {and} \bibinfo{person}{Long Quan}.}
  \bibinfo{year}{2020}\natexlab{}.
\newblock \showarticletitle{Blendedmvs: A large-scale dataset for generalized
  multi-view stereo networks}. In \bibinfo{booktitle}{\emph{Proceedings of the
  IEEE/CVF Conference on Computer Vision and Pattern Recognition}}.
  \bibinfo{pages}{1790--1799}.
\newblock


\bibitem[Yariv et~al\mbox{.}(2021)]%
        {volSDF}
\bibfield{author}{\bibinfo{person}{Lior Yariv}, \bibinfo{person}{Jiatao Gu},
  \bibinfo{person}{Yoni Kasten}, {and} \bibinfo{person}{Yaron Lipman}.}
  \bibinfo{year}{2021}\natexlab{}.
\newblock \showarticletitle{Volume rendering of neural implicit surfaces}.
\newblock \bibinfo{journal}{\emph{Advances in Neural Information Processing
  Systems}}  \bibinfo{volume}{34} (\bibinfo{year}{2021}),
  \bibinfo{pages}{4805--4815}.
\newblock


\bibitem[Zhou and Jacobson(2016)]%
        {thingi10k}
\bibfield{author}{\bibinfo{person}{Qingnan Zhou} {and} \bibinfo{person}{Alec
  Jacobson}.} \bibinfo{year}{2016}\natexlab{}.
\newblock \showarticletitle{Thingi10k: A dataset of 10,000 3d-printing models}.
\newblock \bibinfo{journal}{\emph{arXiv preprint arXiv:1605.04797}}
  (\bibinfo{year}{2016}).
\newblock


\end{thebibliography}

\end{document}